\documentclass[journal=mamobx,manuscript=article]{achemso}

\usepackage[version=3]{mhchem} 
\usepackage[utf8]{inputenc}             
\usepackage[T1]{fontenc}                
\usepackage{soul}                       
\usepackage{float} 
\usepackage{tabularx}        
\usepackage{booktabs}        
\usepackage{multirow}        
\usepackage{threeparttable}  
\usepackage{threeparttablex} 
\usepackage{longtable}       
\DeclareMathOperator{\Tr}{Tr}

\SectionNumbersOn
\mciteErrorOnUnknownfalse
\setkeys{acs}{usetitle = true}

\author{Daniel J. Arismendi-Arrieta}
\email{darismendi@dipc.org}
\affiliation[DIPC]
{Donostia International Physics Center (DIPC), Paseo Manuel de Lardizabal 4,  
E-20018 San Sebastián, Spain}
\author{Angel J. Moreno}
\email{angeljose.moreno@ehu.eus}
\affiliation[CFM]
{Centro de Física de Materiales (CSIC, UPV/EHU) and Materials Physics Center  
MPC, Paseo Manuel de Lardizabal 5, E-20018 San Sebastián, Spain}
\alsoaffiliation[DIPC]
{Donostia International Physics Center (DIPC), Paseo Manuel de Lardizabal 4,  
E-20018 San Sebastián, Spain}
\title[nanogels paper DJAA]
  {Deformability and solvent penetration in soft nanoparticles at 
liquid-liquid interfaces}

\begin{document}
\begin{center}
\end{center}
\begin{abstract}
Soft nanoparticles hold promise as smart emulsifiers due to their high degree of 
deformability, permeability and stimuli responsive properties. By means of 
large-scale simulations we investigate the structural properties
of nanogels at liquid-liquid (A-B) interfaces and the miscibility of the liquids inside
the nanogels, covering the whole range of interfacial strength from
the limit of single-liquid to the case of stiff interfaces. 
To study the role of the internal architecture and deformability of the nanogel we
simulate a realistic disordered and an ideal regular network, for a broad range of
cross-linking degrees. Unlike in previous investigations on liquid miscibility, 
excluded volume interactions are considered for both the monomers and the explicit solvent particles.  
The nanogel permeability is analysed by using an unbiased grid representation that accounts
for the surface fluctuations and adds to the density profiles
the exact number of liquid particles inside the nanogel. 
The better packing efficiency of the regular network leads to  
higher values of the total liquid uptake and the invasive capacity (A-particles in B-side and {\it viceversa}) 
than in the disordered network, though differences vanish in the limit of rigid interfaces.
Uptake and invasion are optimized at a cross-linking degree that depends on the interfacial strength,
tending to $\sim 15 -20$\ for moderate and stiff interfaces.
As the interfacial strength increases, the miscibility inside the nanogel is enhanced by a
factor of up to 5 with respect to the bare interface, with the disordered networks providing a better mixing than their ideal counterparts. 
The emerging scenario reported here provides general guidelines for tuning the shape, uptake, invasive, and mixing capacities of nanogels adsorbed at liquid-liquid interfaces.

\end{abstract}


\section{Introduction}

Crosslinked polymeric particles represent a special class of soft colloids 
with the ability of adapting their dimensions according to the surrounding 
environment or via an external 
stimulus~\cite{Karg_2019_Richtering,Gupta_2015_Richter,Stuart_2010_Minko}. So 
far, very attractive applications derived from nanogels and 
microgels in suspensions or at fluid-fluid interfaces have been proposed such 
as: controlled drug delivery, catalysis, antifouling coatings, cell 
encapsulation and tissue engineering, among others. 
~\cite{Plamper_2017_Richtering,Agrawal_2018_Agrawal,Echeverria_2018_Soares,
Thorne_2011_Snowden} Deformability, especially at 
interfaces, is the most differentiating feature with respect to core-shell 
colloids and truly hard solid particles~\cite{Deshmukh_2015_Duits}. Due to 
their intrinsic stimuli-responsive properties they may prove beneficial in a 
wide range of technologies such as thermocromic components in smart 
windows~\cite{Serpe_2019}, high-performance 
bio-lubricants~\cite{Torres_2018_Sarkar}, controlling 
the metastability of emulsions and 
foams~\cite{Schmitt_2013_Ravaine}, directing self-assembly 
processes~\cite{Yunker_2014_Yodh}, or even as models for translating our 
current understanding of submicron synthetic gels into food based 
systems~\cite{Kwok_2019_Ngai,Murray_2019}.

In the context of the oil-water interface, 
emulsion 
stabilization still remains as a very active research field with novel concepts 
constantly being 
introduced~\cite{Ngai_2014_Bon,Gupta_2016_Doyle,Wu_2018_Striolo}. For 
instance, solid particles 
continue to being explored for understanding their active or passive role in 
the self-fragmentation process of armored droplets~\cite{Sicard_2019_Striolo}, 
while the 
interfacial behavior of more complex particles has been considered only 
recently~\cite{Ngai_2005_Auweter}. Current efforts are focused on designing 
responsive emulsions that stabilize or break on 
demand~\cite{Wiese_2016_Richtering} with the help of smart emulsifiers. 
Therefore, the inherent interfacial activity and stimuli-sensitive properties of 
soft particles could facilitate the 
control and formation of responsive elastic barriers that interfere with 
coalescence and ripening type processes~\cite{Schmitt_2013_Ravaine}. 
However, given the multifactorial character of this problem, studies very often 
rationalize their results by varying one parameter at the time and consequently 
few attempts have been made to characterize such systems in a general 
way~\cite{Schmitt_2013_Ravaine}. 

It is clear that understanding the way such particles deform on adsorption at 
interfaces is very important from the fundamental and applied point of view. 
Nevertheless, our ability to make, measure and modelling these objects is 
nowadays 
the principal limiting factor. With the advances in synthesis methods, it is 
possible to design microscale particles with complex 
architectures~\cite{Richtering_2014_Saunders}. Nevertheless, for the case of 
nanoscale analogous, it is still very challenging to generate tailored internal 
topologies and more importantly to maintain a uniform size distribution. 
Additionally, if internal structure details are required, then they are very 
difficult to be detected by experimental techniques as deformability 
introduces additional degrees of freedom that current measurements hardly 
distinguish. Even with the arrival of super-resolution microscopic 
techniques~\cite{Siemes_2018_Richtering,Karanastasis_2018_Chaitanya}, a 
detailed topological characterization of microgels is 
still scarce, and in the case of nanogels non-existent. Only few experimental 
studies have quantified the location of microgels at liquid-liquid interfaces 
by introducing the concept of two contact angles on lens-like shape 
particles~\cite{Geisel_2012_Richtering, Cristofolini_2018_Isa}. This however 
rises the question if due to the solvent 
penetrability the definition of any contact angle has any sense in such 
systems~\cite{Schmitt_2013_Ravaine,Murray_2019}. Anyhow,
it is evident that alternative approaches are required to 
understand particle conformations at interfaces.

On the other hand, computer simulations allow to correlate such 
aspects in a more controlled way and ultimately generate predictions with 
respect to properties that are hardly reachable by the lack of experimental 
resolution. In doing so two elements have to be considered: i) how to model 
the solvent and ii) how realistic the internal nanoparticle's 
architecture has to be. For the first element, two approaches are often used to deal with 
explicit solvent simulations: molecular dynamics (MD) of particles with excluded volume interactions, and dissipative particle 
dynamics (DPD) with bounded potentials \cite{Karg_2019_Richtering,Martin_2019_Quesada,Kruger_2013_Harting}. 
The latter, as a mesoscopic simulation approach, allows for a direct mapping to Flory-Huggins lattice models and has been 
recently employed to understand swelling or deswelling of microgels at 
suspensions~\cite{Camerin_2018_Zaccarelli}, and their behavior at liquid 
interfaces~\cite{Camerin_2019_Zaccarelli,Rumyantsev_2016_Potemkin,
Gumerov_2016_Potemkim,Gumerov_2017_Potenkim,Gumerov_2019_Potemkim}. The 
former is rarely used due to its huge computational cost, originating from the longer diffusion times 
required with respect to the case of fully penetrable DPD particles. 
Regarding the second element, 
a critical point in current models is that a regular network (e.g. 
diamond lattice) is usually assumed within such soft nanoparticles, whereas with traditional 
synthetic 
approaches a very disordered network abundant in defects (e.g. loops) 
is formed \cite{Moreno_2018_Loverso}. The use of excluded volume interactions between the nanogel's monomers is also
essential in modelling realistic nanogels. Whereas in the case of regular diamond networks
the use of DPD interactions for both the liquid and the monomers \cite{Rumyantsev_2016_Potemkin,
Gumerov_2016_Potemkim,Gumerov_2019_Potemkim} might still be sufficient to model
their spatial distribution, such interactions are clearly inadequate for the monomers in the disordered
networks. An increasing number of permanent contacts (entanglements) between neighboring strands 
is expected by increasing the  degree of cross-linking, or just by flatening of the nanogel in the interface. In such conditions the bounded character of the DPD interactions
will easily lead to violation of the topological constraints through chain crossing. 
Modelling of realistic microgel networks poses a 
computational challenge for developing suitable \textit{in-silico} protocols 
that best 
mimic the internal microgel's structure, density profiles, form factors, as 
well as the kinetics of swelling and deswelling found at different experimental 
conditions. Several efforts to provide realistic models have been reported  
very 
recently~\cite{Gnan_2017_Zaccarelli,Moreno_2018_Loverso,Minina_2019_Kantorovich,Ninarello_2019,Rudyak_2019}.
See, e.g., the recent 
reviews~\cite{Martin_2019_Quesada,Rovigatti_2019_Zaccarelli} for more details 
about the progress 
on modelling gel-like colloidal particles.

In order to elucidate the properties of macroscopic systems such as nanogel 
stabilized emulsions, it is necessary to first understand how isolated 
soft particles behave at interfaces~\cite{Ballard_2019_Bon}. In this work, 
the correlation 
between the internal nanoparticle topology and its influence on deformability, as 
well as solvent penetration at liquid-liquid interfaces is examined by means 
of MD simulations. For the first time this issue is addressed with excluded volume for all the interactions
(liquid-liquid, monomer-monomer and cross-interactions) and the role of the network topology 
(disordered vs. regular) is investigated. Permeability is quantified by an unbiased analysis that
counts the exact number of liquid particles inside the nanogel at every time.
We find that such porous cavities do not exhibit nanoconfined homogeneous mixtures even for low
interfacial strengths. Interestingly enough, the total liquid uptake and invasion (fraction of liquid within the nanogel at the other side
of the interface) are optimized for a cross-linking degree that depends on the interfacial strength. Noticeably, 
the internal topology of the nanogel network (regular or disordered) does not seem to play a major role in the quality of the mixing, except
for stiff interfaces where disordered networks show a higher efficiency. 

The article is 
organized 
as follows. Section 2 describes the simulation details, while the main results 
and discussion are presented in sections 3 and 4. Finally in section 5 the 
conclusions and outlook are presented.
\section{Computational Methodology}
\label{Nanogels synthesis}
In this work, three different simulation stages have been considered, 
see Figure S1 in the supporting information (SI) for an overview of the whole 
workflow. First, the computational synthesis and equilibration of nanogels 
under implicit 
solvent conditions. Within such a stage, two families of homopolymeric nanogels 
were designed: one of them, generally employed in the literature, with a 
regular diamond-like network (from now on named ideal) and a second one with 
an irregular disordered network (from now on named realistic). In the latter case, 
nanogels 
were created with an in-house code following the methodology proposed in 
Refs.~\cite{Loverso_2015_Moreno,Moreno_2018_Loverso}. In  brief, the 
\textit{\textit{in-silico}} 
protocol resembles the design of nanogels, microgels~\cite{Moreno_2018_Loverso} 
and 
globular single-chain nanoparticles~\cite{Loverso_2015_Moreno,Pomposo_2017}, as 
examples of soft 
polymeric objects. In order to investigate the conformations and the effect of 
the internal network structure of the former particles on the
mixing properties of immiscible liquids within nanocavities, we varied the 
fraction of internal crosslink densities, $f_{cl} = N_r/N_m$, with $N_r$ and $N_m$ the number of monovalent cross-linkable groups and the total number of monomers, respectively. This is indeed one of the 
crucial 
factors controlling nanogels behavior. This way, we tunned the elasticity of 
the 
nanogel to adapt itself at different interfacial strengths. In total, 4 types 
of 
nanogels were generated with $f_{cl}=$~3.9, 7.5, 14.0 and 19.9 \%, respectively. 
Realistic nanogels were made of 2000 monomer units, and for the purpose of 
comparison, the ideal ones were as close as possible to the realistic ones; 
with 
2000, 2002, 2021 and 2000 monomer units for each $f_{cl}$ (more details on the 
ideal and realistic nanogels design can be found in 
Refs. 35 and 39 respectively). 
After the synthesis, such 
systems were 
equilibrated for $10^{7}$~ steps 
under Langevin dynamics \cite{AllenBook_2017}. 

\label{Box details}
The second stage consisted of the generation and equilibration of nanogels at 
the interface between two inmiscible liquids with explicit fluid particles. 
For such, we placed the previous equilibrated nanogels at the  center of the 
simulation box, and filled the box with two different liquids (A and B) of monoatomic molecules. Periodic 
boundary conditions were applied in all directions and the liquid/liquid 
interface was set in the $x-y$ plane and perpendicular to the $z$-direction. Nanogel's 
sizes were rationalized to have a 
good compromise between the large computational effort needed to include
explicit solvent 
particles 
and the need of simulating large enough nanogels with a reasonable amount of 
solvent particles trapped in their 
interior. The PACKMOL package~\cite{Packmol_2009} was employed for generating 
the 
initial configuration of the system. Inside the cubic simulation box of sides 
$L 
= 100\sigma$, the two different fluid particles were packed with a particle 
number density $\rho^{\ast} \approx 0.952/\sigma^{3}$~(475000 particles for each liquid 
component + the nanogel). Finally, in order to provide reference conformations 
for 
comparisons at different environments (implicit solvent, as well as explicit 
solvent with monomer-solvent excluded volume interactions), at the 
third stage we make equal the cross-terms interactions between the liquids to 
simulate the case of a nanogel dispersed in an homogeneous mixture (single liquid). 
\label{General simulation details}
For the last two stages, systems were equilibrated for $10^{9}$~ 
steps. Through this paper, all simulations with explicit solvent were carried 
out with the GROMACS 4.6.5~\cite{Gromacs_2015}~package under Newtonian 
dynamics 
in the NVT ensemble. Lennard Jones reduced units are used,  and all 
particle masses, energy and length scales are set as $m = \epsilon = \sigma = 
1$.  The 
reduced temperature and timescales, $T^{\ast}$ an $\tau$, are given by 
$T^{\ast} 
= \epsilon /k_B = 1$~(with $k_B$~the Boltzmann constant) and $\tau = (m\sigma 
^{2})^{1/2} $, respectively. The time step in all stages was $\delta t = 
0.003~\tau$, while the relaxation time for the Nos\'{e}-Hoover 
thermostat \cite{Nose_1984,Hoover_1985} was set to $0.5\tau$.

\label{Model Interactions}
Concerning the interactions, these simulations employ the Kremer-Grest 
bead-spring model~\cite{Kremer_1990_Grest}. All non-bonded 
interactions between beads (nanogel's monomers and liquid particles) are given by the Weeks-Chandler-Anderson (WCA) 
potential~\cite{Weeks_1971_Andersen} to account for excluded-volume 
interactions:

\begin{equation}
U^{WCA}(r) = 4\epsilon\left[ 
\left(\frac{\sigma}{r}\right)^{-12} - 
\left(\frac{\sigma}{r}\right)^{-6} + \left(\frac{1}{4}\right) \right] 
\text{for} \quad r < 2^{1/6}\sigma
\end{equation}

\noindent and zero for $r \geq 2^{1/6}\sigma$, while bonded nanogel's beads 
interact 
through the finite-extensible-non-linear 
spring (FENE) potential given by

\begin{equation}
U^{FENE}(r) = -\frac{1}{2}K_{F} R_{0}^{2}~\text{ln} \left[1- 
\left(\frac{r}{R_{0}}\right)^{2} \right]\text{for} \quad r < R_0 
\end{equation}

\noindent and $\infty$~ for $r \geq R_0$, with $K_{F} = 15\sigma^{-2}$~ and 
$R_{0} = 
1.5\sigma$. 
Thus, for dealing with a finite polymer network with $m$~monomers, within a 
two-component mixture of explicit solvent molecules A and B, the 
interaction energy of the system is written as
\begin{equation}
U = \sum_{a,b,m=1}^{N_a,N_b,N_m}U^{WCA} + 
\sum_{bonds}U^{FENE}
\end{equation}

\noindent where the purely repulsive nature of the WCA potential 
mimics good solvent conditions between the solvent and nanogel's monomers, 
while the combination of FENE and WCA guarantees chain uncrossability by 
limiting the fluctuation of bonds in the nanogel. Additionally, we have 
considered the nanogel having the same affinity for both liquid 
components, $U^{WCA}(r_{ma}) =  
U^{WCA}(r_{mb}) = U^{WCA}(r_{mm})$. Such an 
assumption,  very hard to implement in experiments,  provides 
however a quasi-ideal scenario 
for investigating the effect of the network topology on the solvents' mixing 
inside the nanogel. Nonetheless, to account for the overall effect of 
interfacial fluctuations, 3 interaction strengths have been chosen for the liquid-liquid interactions
($\epsilon_{AB}= 3, 5~\text{and}~15$), while $\epsilon_{AB}=1$ 
is considered the reference case for the homogeneous mixture (single liquid).  
\section{Results}
\subsection{Nanogel's deformability} 
In order to quantify the average nanogel's shape and size under different 
environments, the gyration tensor in the laboratory frame is employed and 
defined as
\begin{equation}
\c{G}_{\mu \nu}= \frac{1}{2N_m^2} 
\sum_{i=1}^{N_m}\sum_{j=1}^{N_m}(r^{(i)}_\mu-r^{(j)}_\mu)(r^{(i)}_\nu-r^{(j)}
_\nu)
\end{equation}
where the summation is performed over the $N_m$~monomer beads in the nanogel, with 
$\mu,\nu\in \{x, y, z\}$ as the Cartesian coordinates of the 
$i^{th},j^{th}$ beads.  Additionally, $\c{G}_{\mu 
\nu}$ is also decomposed into perpendicular ($\c{G}_{\perp}=\c{G}_{zz}$) and 
parallel 
 ($\c{G}_{\parallel}=\c{G}_{xx}+\c{G}_{yy}$) components to obtain the size of 
the nanogels relative to their orientation at the interface. To describe the 
overall spatial distribution of monomers in the nanogels $\c{G}_{\mu \nu}$ has 
to be diagonalized. After that, structural features are extracted by 
combining the corresponding eigenvalues 
($\lambda_{1}~\geq~\lambda_{2}~\geq~\lambda_{3}$) into a group of shape 
parameters~\cite{Rawdon_2008_Millett,Rudnick_1987_Gaspari}. In this work, three 
of them are employed: 
The squared radius of gyration (${R_g}^{2}$),

\begin{equation}
\label{Rg}
\Tr \c{G}_{\mu \nu}=\c{G}_{\perp}+\c{G}_{\parallel}={R_g}^{2}=\lambda_{1} + 
\lambda_{2} + \lambda_{3} 
\end{equation}
as the sum of the principal moments. The asphericity ($a$)
\begin{equation}
\label{asphericity}
a=\frac{(\lambda_{2} - \lambda_{1})^2 + (\lambda_{3} - \lambda_{1})^2 + 
(\lambda_{3} - \lambda_{2})^2}{2(\lambda_{1} + \lambda_{2} + \lambda_{3})}
\end{equation}
$0~\leq~a~\leq~1$, zero for symmetrical conformations (e.g. a sphere, 
a cube) 
and one meaning objects with cylindrical symmetry (e.g. a rod). The 
prolateness ($p$),
\begin{equation}
\label{prolateness}
p=\frac{(3\lambda_{1} - {R_g}^2)(3\lambda_{2} - 
{R_g}^2)(3\lambda_{3} - {R_g}^{2})}{2(\lambda_{1}^2 + \lambda_{2}^2 + 
\lambda_{3}^2 - 
\lambda_{1}\lambda_{2} - \lambda_{1}\lambda_{3} - \lambda_{2}\lambda_{3})^{3/2}}
 \end{equation}
$-1~\leq~p~\leq~1$, quantifies deviations between perfectly oblate ($p=-1$) 
and prolate ($p=1$) objects.

\begin{figure}[p]
\centering\includegraphics[scale=1]{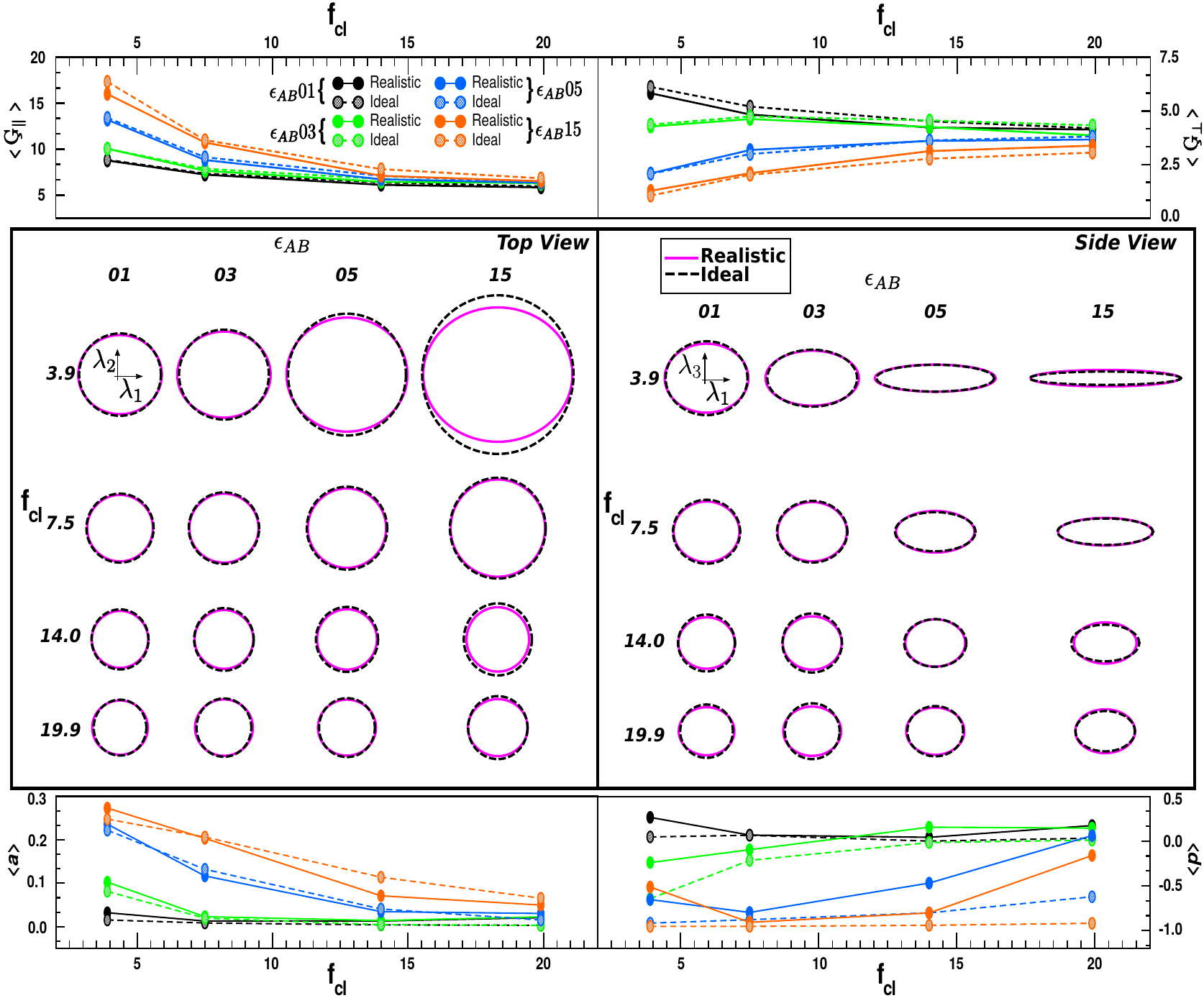}
\caption[shape parameters]{Shape parameters as a function of cross-link 
densities for different liquid compatibilities in realistic (continuous lines 
with solid circles) and ideal (dashed lines with textured circles) nanogels. 
Top 
panels represent the parallel (left) and perpendicular (right) decomposition of 
the radius of gyration. Middle panels correspond to the top (left) and side 
(right) view of the gyration ellipsoid (eigenvalues). Bottom panels correspond 
to the 
asphericity (left) and prolateness (right).}
\label{fig-shapes}
\end{figure}

In Figure \ref{fig-shapes} the main trends for the realistic 
and ideal nanogel are presented as a function of the degree of 
cross-linking and liquids compatibility, while Table S2 in the SI 
summarizes the average shape 
descriptors of nanogels under different environments. The radius of 
gyration has been decomposed into parallel and perpendicular components
to quantify the nanogel's size (see top panels in Figure \ref{fig-shapes}). 
The absence of explicit solvent leads to bigger nanogel's sizes (about 20\%) than when it 
is embeded in the dense liquid, showing the relevance of excluded volume interactions even in the 
absence of the adsorbing interface.
As the cross-link density increases, the nanogel's size decreases monotonically in the $xy$-plane 
and converges from above the $\c{G}_{\parallel}$~value of the homogeneous 
mixture case. No big differences are observed between the nanogels of both topologies and, except in the case
$f_{cl} =3.9$, the cross-linking degree has only a minor effect on the nanogel size when it is confined 
by the weakest interface ($\epsilon_{AB}=03$) or solvated in the 
homogeneous mixture ($\epsilon_{AB}=01$).
Only for strong interfaces ($\epsilon_{AB}\geq 05$) the cross-linking degree plays a major role.
In the $z$-direction, the 
nanogels show the opposite behavior and their size converges from below the 
$\c{G}_{\perp}$~ of the homogeneous mixture case. For the weakest interface, 
the nanogel's perpendicular size 
remains almost constant while for $\epsilon_{AB}\geq 05$ it grows as the degree of cross-linking 
increases. Only when the degree of cross-linking is high the liquid compatibility has a minor effect
on $\c{G}_{\perp}$~and 
$\c{G}_{\parallel}$, otherwise nanogels deform according to how flexible their 
internal networks are. 
 
The fact that nanogels can adapt their shapes to the confining environment is 
quantified by the asphericity and prolateness parameters. In particular, for 
the asphericity (see bottom-left panel in Figure \ref{fig-shapes}), there is a 
clear trend toward spherical objects ($a \rightarrow 0$) as the degree of cross-linking grows. This 
indicates that nanogels are not able to deform anymore, no matter the strength 
of the interface. As the degree of cross-linking decreases and the interface 
becomes more rigid, nanogels lose the typical spherical shape found in the 
homogeneous mixture. The prolateness of such spheroidal objects (see 
bottom-right panel in Figure \ref{fig-shapes}) describes how elongated (prolate) 
or flatened (oblate) the nanogels are when 
adsorbed at the interface or inmersed in the single liquid. For the 
latter, nanogels shows a prolateness close to zero as expected. Nevertheles, 
due to the inhomogeneous nature of realistic nanogel networks, they tend to 
have a more prolate character than their ideal counterparts. At the interface between two inmiscible liquids, 
nanogels gradually flaten as the strength of the interface increases and the 
degree of cross-linking decreases. For realistic nanogels, we again observe 
more elongated objects than for the regular networks as the number of cross-links and the interfacial 
strength increases. Due to the sensitivity of the prolateness, it seems that 
realistic nanogels are very different from their ideal counterparts, however if 
we plot the top and side view 
projections of the gyration tensor eigenvalues (see middle panels in Figure 
\ref{fig-shapes}), such differences are less pronounced and now it is easier to 
appreciate the deformability of such soft objects. Overall, no significant 
differences are observed between realistic and ideal nanogels, except in the case 
of $f_{cl}=$ 3.9 and $\epsilon_{AB}=$15, which just indicates that the realistic 
nanogel is less deformable in the direction of the interfacial plane than 
its  ideal counterpart. 
Figure S3 in the SI shows the 
parallel and perpendicular 
swelling curves, for the interfacial case against the nanogel
dispersed in the homogeneous mixture. Clearly the most deformable nanogels (low $f_{cl}$)
show the stronger dependence of their size on the interfacial strength.

\begin{figure}[p]
\includegraphics[width=16cm]{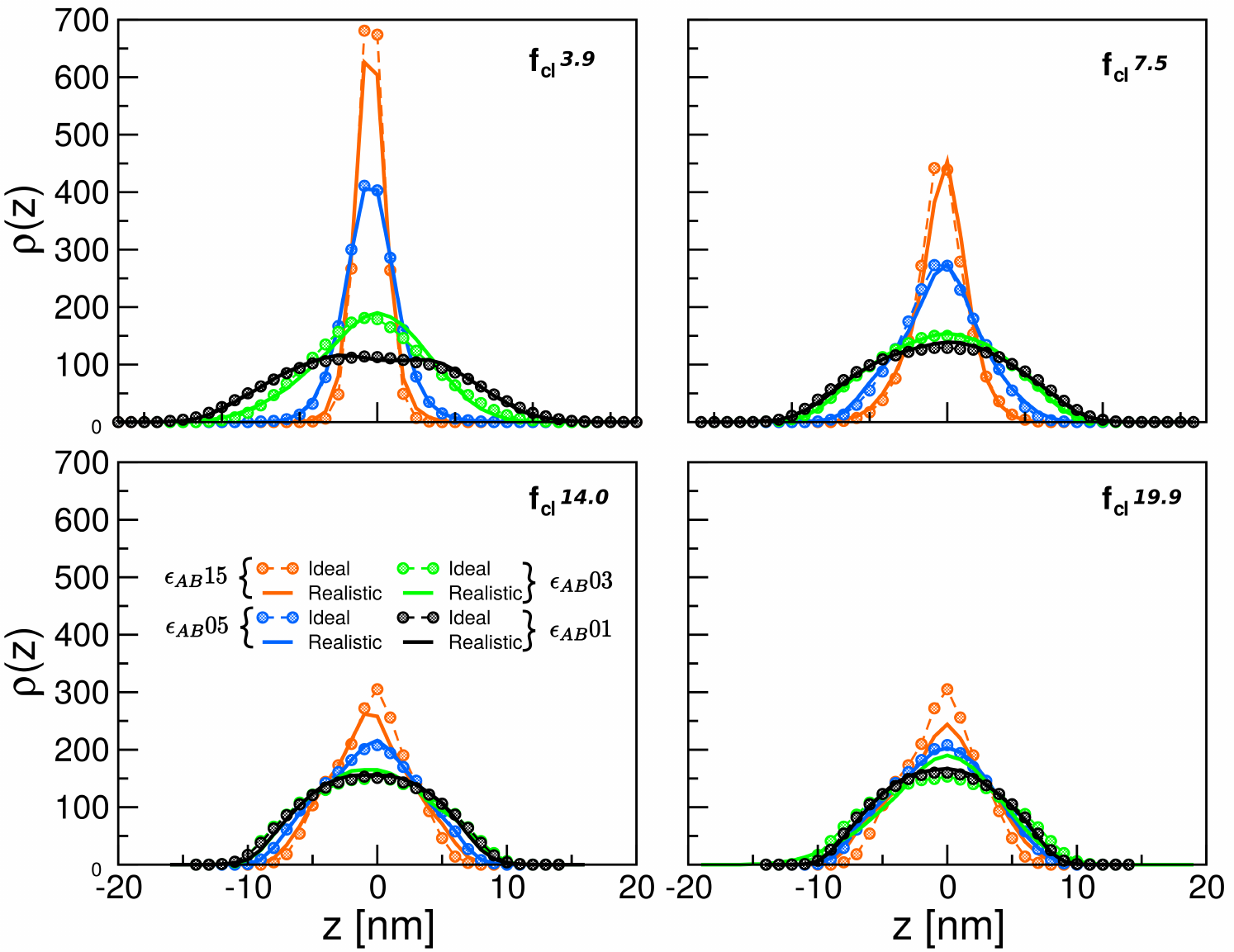}
\caption[density profiles polymers]{Density profiles of 
realistic (continuous lines) and ideal (dashed lines with textured circles) 
nanogels with $f_{cl}=$ 3.9, 7.5, 14.0 and 19.9, taken 
orthogonally to the $xy$~plane for different degrees 
of liquid's affinity: $\epsilon_{AB}=01$ corresponds to the single 
liquid, while $\epsilon_{AB}=$03, 05 and 15 represent liquid mixtures with 
different degree of incompatibility. All data are normalized so that the area 
below the curve corresponds to the total number of particles in the nanogel.}
\label{fig-density_profiles}
\end{figure}

More information on how the nanogels deform at the interface is gained by 
examining the average number density profiles $\rho(z)$~reported in Figure 
\ref{fig-density_profiles}. 
In the case of the incompatible liquids, such profiles measure the amount of 
nanogel monomers at a distance $z$ from the interface, while for the case of 
nanogels immersed in the 
single liquid, such curves correspond to the number of nanogel monomers along 
the z coordinate in the simulation box. For the purpose of comparison, the 
homogeneous mixture 
will serve as a reference to highlight the differences between flattening or 
compression of nanogels when adsorbed at the interface.

For the nanogel dispersed in the single liquid, the distribution narrows as the 
degree of cross-linking increases. At the interface, 
nanogels' deformability comes from the interplay between the interfacial 
tension forces and the amount of structural nodes inside the nanogel. As 
expected, the maximum of $\rho(z)$ is found for $z=0$~since nanogel 
monomers tend to 
reside at the interface in order to avoid direct contacts between the 
incompatible liquids. As aforementioned, the 
inhomogenous distribution of cross-links in realistic nanogels does not seem to significantly
affect their overall shape  when compared to their ideal 
counterparts. This also seems to be the case in terms of $\rho(z)$. However for 
very strong interfaces ($\epsilon_{AB}=15$) the ideal nanogels generally 
accumulate more monomers in the interface ($z=0$) than realistic ones. The shape of such 
profiles is strongly correlated with the side view projections presented in the 
middle-right panel of Figure \ref{fig-shapes}. Further, there is a clear 
transition on how the nanogel monomers organize from the single liquid to the 
case of two inmiscible liquids. The central peaks decrease in 
intensity and broaden indicating that nanogels are less deformable and occupy 
more volume in the $z$-direction as the degree of cross-links increases. For very 
low degree of cross-linking the distributions narrow sharply as the 
interfacial strength increases, thus reaching a narrow peak that resembles a 
single layer of nanogel monomers. 

\subsection{Solvent penetration in deformable porous cavities}
There are several ways to calculate how many solvent particles 
reside inside nanogels at each time frame. One 
might think that just using the components of the 
radius of gyration or the main ellipsoid axis to select the particles inside 
such boundaries would be sufficient. However such criteria 
might bias the interpretation for various reasons: (i) how one counts the 
liquid particles will intrinsically depend on how one defines the enclosing surface 
that best describes the shape of the nanogel; (ii) even if in average the 
nanogel has an specific shape, it is important to pick up a geometric 
enclosing surface that also accounts for the instantaneous conformational 
fluctuations of the nanogel; (iii) and most importantly, the boundaries of such 
geometrical constructions correspond to an enclosing smooth surface and do not 
account for the nanogel protrusions of the outer layers. 
An alternative is to 
create a grid representation of the nanogel to count the number of 
solvent beads inside it. In particular, for each configuration we first construct an
enclosing parallelogram surrounding the nanogel. The sides in all directions are selected so that the parallelogram encloses
all the nanogel's monomers. Then we divide the 
parallelogram with a grid mesh size of $\Delta=\sigma$. For each column $(x,y)$ in the 
grid we look for the highest and lowest nanogel's monomer in the $+z$ and $-z$ directions, respectively. 
During the panning, for each column $(x,y)$ we count the cells with liquid beads (A or B) or nanogel monomers that are
enclosed by the highest and lowest nanogel's monomers at that $(x,y)$ (both monomers are also counted). In this way the counting 
considers the nanogel internal topology fluctuations, and does not count particles out of the nanogel's outer surface 
neither miss particles inside it ---an artifact that emerges when a static smooth enclosing  surface is used as, e.g., a sphere of
radius $R_{\rm g} = \langle R_{\rm g}^2 \rangle^{1/2}$. Indeed our method, by construction, counts at every time the exact 
number of liquid particles that are inside the nanogel.
In Figure S4, an instantaneous side view of the realistic nanogels is presented 
for the purpose of illustrating how the liquids reside inside the pores of 
such nanocavities. Additionally, two movies (S5 and S6 in the SI) show the side 
and top view projections of realistic and ideal nanogels with the solvent from 
bulk phases residing inside them.

\begin{figure}[p]
\includegraphics[width=12.3cm]{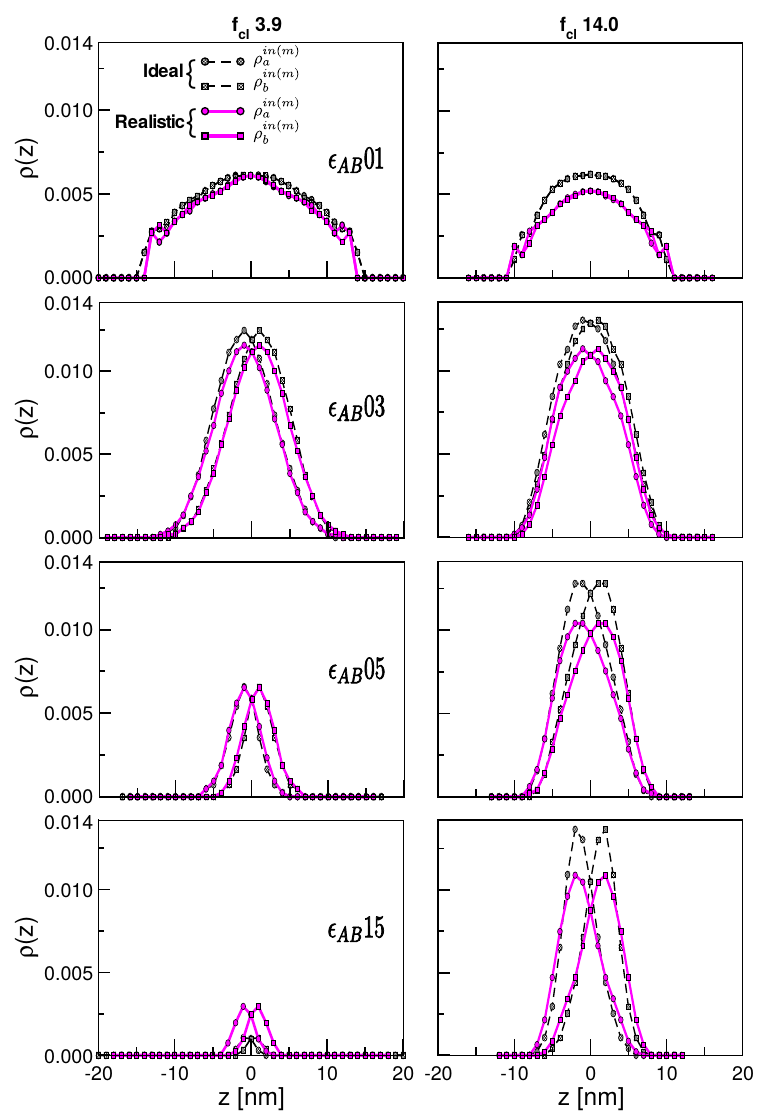}
\caption[density profile of liquids inside nanogels]{Density profile of 
solvent beads inside realistic and ideal nanogels. 
Liquids A and B 
correspond to $\rho_{a}^{in
(m)}=\left<\frac{N_a^{in (m)}}{N_a^{in (m)}+N_b^{in (m)}+N_{m}}\right>$~and 
$\rho_{b}^{in
(m)}=\left<\frac{N_b^{in (m)}}{N_a^{in (m)}+N_b^{in (m)}+N_{m}}\right>$
respectively. $N_{a,b}^{in (m)}$ and $N_{m}$ denote the number of liquid (A or B) particles and nanogel monomers, respectively,
inside the nanogel according to the grid representation. For each $z$-value all the
counts of the different $(x,y)$-columns are summed, and averages are made over the 
different time frames.
Curves are presented 
for the single liquid case as reference for comparison (upper panels), 
while the rest of the panels show the transition from the weakest to 
the strongest interfacial strength. The effects of the nanoparticle 
deformability are addressed from left to right, with $f_{cl}=$3.9 
and $f_{cl}=$14.0 as examples of the lowest and intermediate cases of flexible 
nanocavities.}
\label{fig-mixing}
\end{figure}

Figure~\ref{fig-mixing} shows the average relative amount of liquid 
particles A and B inside the nanogels, $\rho_{a}^{in (m)}$~and $\rho_{b}^{in 
(m)}$ respectively, in the perpendicular direction to the interface, while for 
the single liquid case profiles correspond to the $z$-axis of the simulation box. 
Two fractions of cross-links, 3.9 and 14.0 are presented here, while the 
remaining ($f_{cl}=$7.5 and $f_{cl}=$19.9) are included in Figure S7 of the SI.
To account for specific differences introduced by employing a homogenous or 
inhomogeneous distribution of nodes in the nanogel's internal network, curves 
have been normalized by the total sum of the nanogel's monomers and the 
liquids' particles inside such deformable objects.
Their maxima do not lie at the center of the interface. Full overlap 
between $\rho_{a}^{in (m)}$~and $\rho_{b}^{in 
(m)}$ only occurs for the homogeneous mixture (upper 
panels). As the incompatibility between the liquids increases, the peak 
maxima separate from the interface and the liquids are also less miscible 
inside the nanogels. 
Additionally, the amount of liquids that the nanogel can retain for the case of highly 
incompatible liquids decreases considerably at low fractions of 
cross-links ($f_{cl}=$3.9), while for the rest of cross-linking degrees the curves 
reach a maximum of 0.014 for $f_{cl}=$7.5 (see Figure S7 of the SI) or 
remain almost constant as in the case of $f_{cl}=$14.0.

\begin{figure}[p]
\includegraphics[width=12cm]{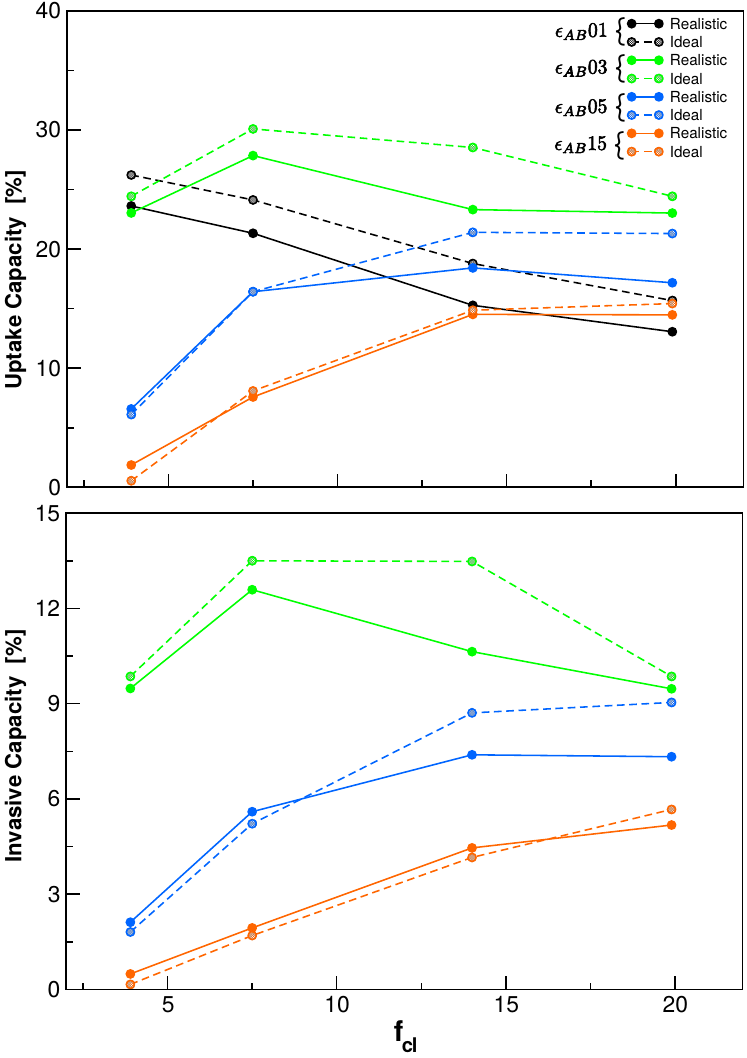}
\caption[capacities]{Solvent uptake $\left<\frac{N_a^{in (m)}+N_b^{in 
(m)}}{N_a^{in (m)}+N_b^{in 
(m)}+N_{m}}\right>$ and invasive $\left<\frac{N_a^{in (B_m)}+N_b^{in 
(A_m)}}{N_a^{in (m)}+N_b^{in 
(m)}+N_{m}}\right>$ capacities of 
liquids inside the realistic and ideal nanogels as a function 
of the particle deformability $f_{cl}=$~3.9, 7.5, 14.0 and 19.9, as well as the 
whole range of liquid compatibilities employed in this study. $N_\alpha^{in (\beta_m)}$
denotes the `invasive' $\alpha$-liquid particles, i.e., those inside the nanogel that are at the $\beta$-side of the interface.}
\label{fig-capacities}
\end{figure}

By comparing the differences between realistic 
and ideal nanogels, it is observed that a more homogeneous distribution of 
nodes in the internal structure leads immediately to an increase in nanogel's 
capacity (higher maxima). This is due to the fact that a regular node 
arrangement enhances the packing and distribute the liquids in a more 
efficient way. Such effects can be appreciated more clearly in 
Figure~\ref{fig-capacities}, where the uptake and invasive capacities of 
nanogels are represented, top and bottom panels respectively. The uptake is 
quantified by adding the areas below the curves in Figures 
\ref{fig-mixing} and S7. The invasive capacity is obtained by summing 
the areas below the $\rho_{a}^{in (m)}$ in the B-side ($\rho_{a}^{in (m_B)}$) 
and $\rho_{b}^{in (m)}$ in the A-side ($\rho_{b}^{in (m_A)}$), respectively, i.e., the
liquid within the nanogel that is at the other side of the interface. For 
the single liquid case, the uptake capacity decreases in a monotonous way as the 
fraction of cross-links increases for both the realistic and ideal nanogels. 
This is an expected result as the nanogels become less deformable and therefore 
cannot swell to enclose more liquids inside them. For the immiscible liquids 
the trend shows a different behavior. For the weakest interface $\epsilon_{AB}= 03$
the uptake is weakly and non-monotonically dependent of the cross-linking degree, with a maximum at
$f_{cl} \sim 5$. For higher
interfacial strength, as the network becomes less deformable (higher $f_{cl}$) 
the uptake capacity increases until reaching a plateau where nanogels cannot 
accommodate more liquids, no matter the degree of cross-linking. It is obviously expected 
that if the number of nodes in the network further increases, the capacity to capture 
solvent inside their cavities will drop at some point, until reaching the limit case of a rigid 
nanoparticle (no uptake). In terms of the total mass (nanogel monomers and the 
liquids inside them), we observe that the invasive capacity (A-liquid in the B-side and {\it viceversa}) 
follows a trend 
that is very similar to the uptake capacity. Indeed it appears that for the 
weakest interface almost half of the solvent particles inside the nanogel are residing on the 
other side of the interface. 

\begin{figure}[p]
\includegraphics[width=12cm]{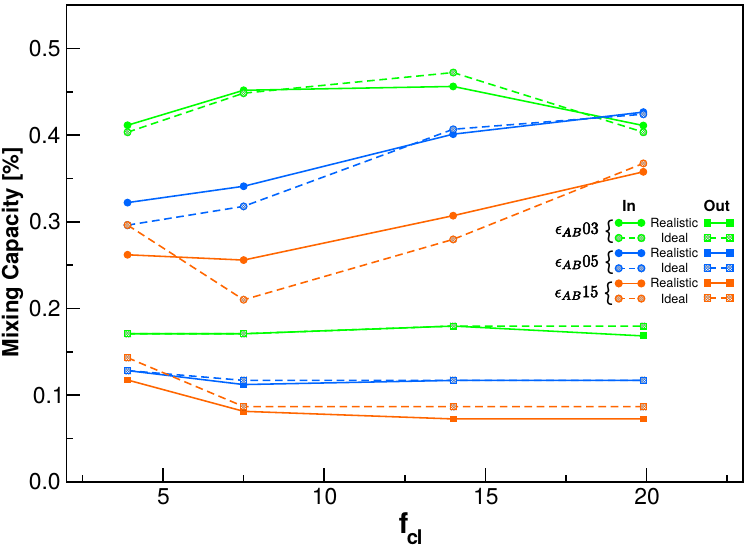}
\caption[quality of mixing]{Quality of mixing $\left<\frac{N_a^{in 
(m_B)}+N_b^{in 
(m_A)}}{N_a^{in (m)}+N_b^{in 
(m)}}\right>$ inside and outside of realistic 
and ideal nanogels as a function of particle deformability $f_{cl}=$~3.9, 7.5, 
14.0 and 19.9, and for the whole range of interfacial strengths.}
\label{fig-qofmixing}
\end{figure}

\begin{figure}[p]
\includegraphics[width=12cm]{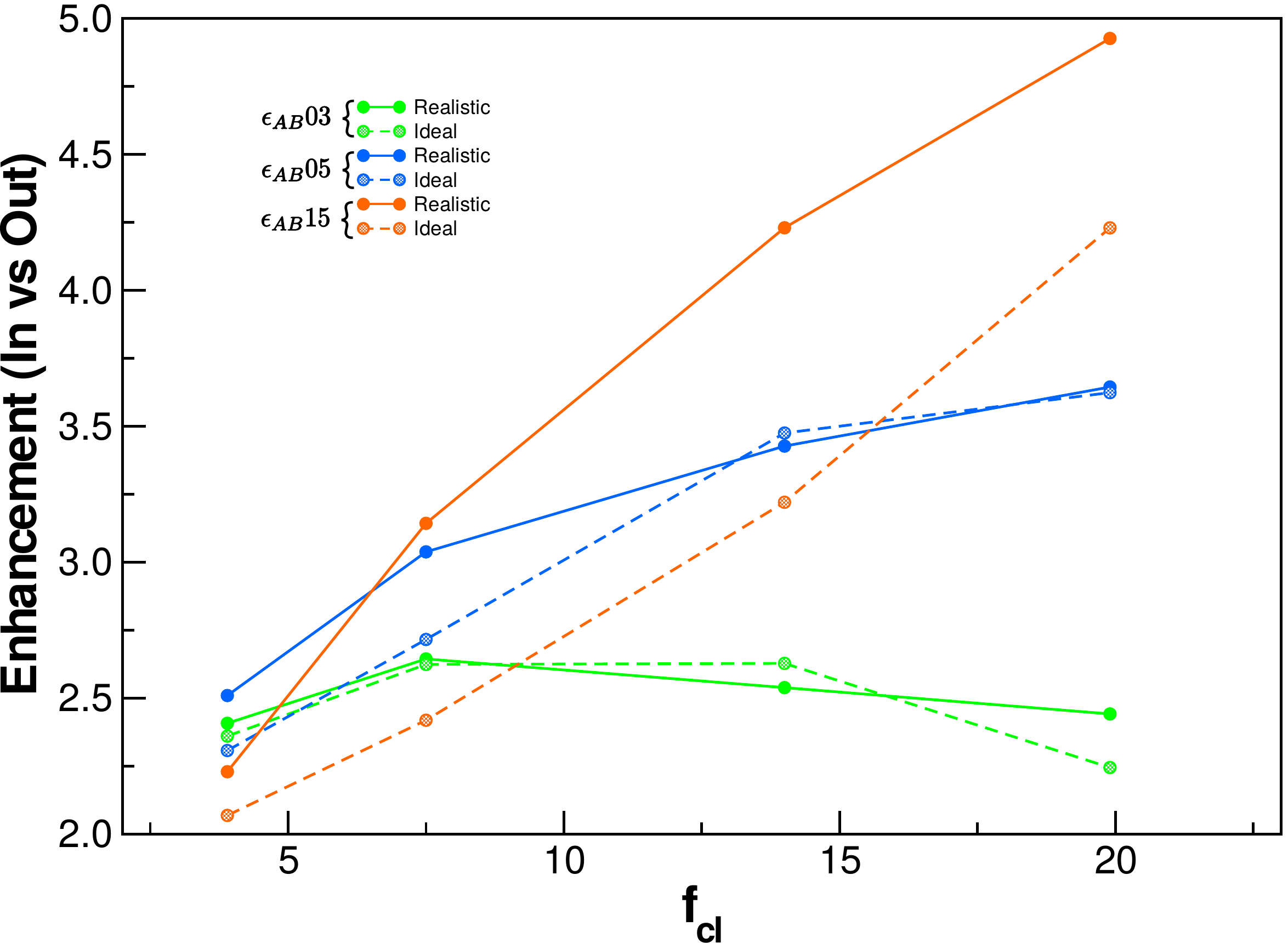}
\caption[enhancement]{Miscibility enhancement for realistic and ideal 
nanogels as a function of the density of cross-links. Such curves correspond to 
the ratios between the quality of mixing inside and outside the nanogels, and they 
are represented for the whole range of interfacial strengths}
\label{fig-enhancement}
\end{figure}

Given such a strong correlation, in Figure 
\ref{fig-qofmixing} we present the ratio between the invasive and uptake 
capacity to quantify the quality of mixing (i.e., which fraction of the total liquid captured by the nanogel is at the
other side of the interface). Moreover, we also include the 
mixing capacity outside the nanogel, i.e., the  fraction of interfacial liquid at the other side of the bare interface.
This was done to verify if miscibility is 
really enhanced inside the cavities of such deformable particles. 
A simulation at the same number density but with no nanogel was carried out for calculating the mixing capacity
in the bare interface. For the sake 
of a fair comparison, 
the integration limits in the bare interface
were taken as those of the average nanogel size in the $\pm z$ directions. 
For all the interfacial strengths we observe a clear enhancement of the 
miscibility of solvent particles inside the nanogel with respect to that 
in the bare interface (see Figure \ref{fig-enhancement}). For the 
weakest interface, such enhancement ($\approx$ 2.5 factor) is almost independent 
on the degree of cross-linking, while for the other 
interfaces the miscibility inside the nanogel increases considerably from a 
factor of 2-2.5 for both the intermediate and rigid interfaces at low 
$f_{cl}=$3.9, to almost a factor of 3.5-5 at high degree of cross-linking
$f_{cl}=$19.9.  By comparing the mixing capacity between realistic and ideal 
nanogels, we observe that for the weakest interface there is no significant 
differences between them, while for the  rigid interface 
the realistic nanogel shows a  considerably better capacity of mixing.

\section{Discussion}
Before continuing let us recall that up to date the focus of simulations and 
experiments has been oriented towards explaining structural properties of 
microgels in suspensions or at interfaces. Only a few recent works have addressed the question
of the liquids' miscibility inside the microgel 
\cite{Gumerov_2016_Potemkim,Rumyantsev_2016_Potemkin,Gumerov_2017_Potenkim,Gumerov_2019_Potemkim}. 
The influence of the network 
topology in features such as softness, particle permeability and their 
correlation is discussed here for large-scale simulations with explicit solvent 
molecules and excluded volume interactions. Previous experimental 
studies~\cite{Schmitt_2013_Ravaine,Destribats_2014_Schmitt, 
Keal_2017_Monteux,Tatry_2019_Schmitt} show that the resistance or 
fragility of emulsions generated with analogous particles in the micro scale 
depends on many factors such as: (i) the cross-link density, 
(ii) the size, (iii) the arrangement and packaging, as well as (iv) the 
processing or emulsification pathway.
All of them from one way or another 
apparently modulate deformability, which is the key factor that 
differentiates soft from hard particles.
Nevertheless, the impact on 
interfacial properties is not clear yet as there are some discrepancies 
concerning if deformability really facilitates adsorption and consequently 
enhance emulsion 
stability~\cite{Murray_2019,Schmitt_2013_Ravaine,Destribats_2011_Schmitt}. 
Ref.~\cite{Destribats_2014_Schmitt} has reported that if the size of microgels 
is increased for 
a fixed cross-link density, this immediately leads to a decrease in surface 
coverage, in turn deteriorating emulsion stability. Under the former 
assumption, small microgels or nanogels could offer additional advantages such 
as a better bridging inhibition, films that withstand 
mechanical stress and ultimately more handleable emulsions.

In this work, deformability was accessed by changing the density 
of cross-links. It has been demonstrated that it is the governing 
parameter in emulsions stabilized with small 
microgels~\cite{Destribats_2014_Schmitt}. Their morphology can also be 
manipulated through different interfacial strengths. For instance, highly 
deformed particles were found in our simulations for nanogels adsorbed at very 
rigid interfaces. Such conformations are analogous to those observed in 
emulsions under high shear rates~\cite{Destribats_2013_Ravaine}. 
Despite that our nanogels have the same affinity for both liquids, we 
still observe their characteristic broadening at the interface as the 
degree of cross-links increases. Indeed, our polymer density profiles correlate 
very well with the results reported for microgels with fried-egg-like 
shapes~\cite{Destribats_2014_Schmitt,Camerin_2018_Zaccarelli,
Camerin_2019_Zaccarelli,Harrer_2019_Vogel}. Even though these simulations 
correspond to nanogels that are far from the overlap concentration, such 
deformable objects are intrinsically permeable. This implies that droplet 
adhesion can be facilitated or inhibited by tunning the invasive capacity 
of such particles. As we saw in the previous section, the invasive capacity is 
reduced for highly deformable nanogels. In this context, nanogels have an 
additional advantage with respect to their counterparts in the microscale as 
their invasive capacity will depend on the object size and shape ---the extent of invasion is limited to distances much smaller than
typical microgel sizes, except close to the limit of miscibility. For the case 
of densely packed nanogels, we expect that this quantity will have few 
variations and it might help to explain why single nanogel monolayers are more 
adhesive than emulsion droplets made with thicker 
barriers~\cite{Keal_2017_Monteux}.

Regarding solvent penetrability, particularly in the context of mixing, 
Potenkim {\it et al.} \cite{Rumyantsev_2016_Potemkin,Gumerov_2016_Potemkim,Gumerov_2019_Potemkim} 
showed that liquid 
miscibility is higher within ideal microgels than in the bulk, and that liquids 
inside them can be in a partial or fully mixed state. The results presented in 
this study go in the same direction, however there are noticeable differences 
between the liquid density profiles presented in this work, and those reported 
by Refs.  
\cite{Rumyantsev_2016_Potemkin,Gumerov_2016_Potemkim,Gumerov_2019_Potemkim}. In particular, no homogeneous mixing is 
observed inside the nanogels for the whole range of cross-links and interfacial 
strengths employed in this study (in contrast to the flat profiles observed in such references for low or moderate interfacial strength). 
This indicates, first, that smooth geometrical enclosing 
volumes \cite{Rumyantsev_2016_Potemkin,Gumerov_2016_Potemkim,Gumerov_2019_Potemkim} are not the best choice to 
discriminate solvent particles inside porous nanocavities, since they do not capture
the fluctuations of the microgel's surface and easily count for liquid particles that are close but out of the microgel.
The grid representation approach presented in this work offers a  
suitable and unbiased choice to select solvent particles inside soft nanoparticles with strong 
deformability. How the solvent particles are modeled is another source 
of discrepancies. In fact, the introduction of explicit solvent particles with 
excluded volume interactions immediately rules out the possibility of liquid 
overlapping in the simulation cell and, more importantly, inside the nanogel's cavities. 
Modelling the solvent with a dissipative particle potential 
facilitates analogies with Flory-Huggins lattice models \cite{Gumerov_2016_Potemkim}. However care must be 
taken as the 
ultrasoft and bound character of such a potential, and the high density of DPD particles
inherent to the method, allow for liquid particle overlapping. This is 
especially critical inside the polymeric nanoparticles, where 
bulk solvent can permeate more easily due to screening of non-favorable contacts 
via the polymer monomers. DPD interactions have aso been used for the monomer-monomer interactions
in the diamond networks of Refs.~\cite{Rumyantsev_2016_Potemkin,Gumerov_2016_Potemkim,Gumerov_2019_Potemkim}.
However, the use of excluded volume for the monomer-monomer interactions
is essential to investigate the nanogels, at least for disordered networks, since in these systems bounded DPD interactions will easily
lead to violation of topological constraints (entanglements) through chain crossing, especially when
the nanogel flatens at the interface.
Finally, if we think about the specific role of 
network topology (ideal vs realistic nanogels), caution is advised particularly 
from the experimental point of view as the community is rapidly transitioning 
from simple, approximately homogeneous gel particles towards more sophisticated 
realizations, e.g. in terms of chemistry, architecture and 
softness~\cite{Richtering_2014_Saunders,Gumerov_2019_Potemkim}. As our results 
indicate, network topology does not seem to critically influence the 
quality of mixing. Our simulations show that realistic disordered nanogels even have, 
for intermediate and very rigid interfaces, slightly better mixing capacities than ideal regular networks. Therefore rushing to 
enhance their efficiency by tunning the network topology should be seen with some
skepticism. From another perspective, even if the internal nanogel topology 
does not critically influence mixing (realistic vs ideal), we believe that this will be an 
important factor in emulsion stability. For instance, it would be 
interesting to push forward more complex nanoparticle architectures to examine if 
enhanced mechanical 
properties are observed in the densely packed responsive monolayers of nanogel 
or microgel stabilized 
emulsions~\cite{Maestro_2019,Richtering_2014_Saunders,Deshmukh_2015_Duits}. 

\section{Conclusions}
We have presented a systematic and intensive computational study to 
understand the conformational properties of a soft colloidal 
particle (nanogel) adsorbed at the interface between two inmiscible liquids. We 
have synthetized realistic and ideal nanogels \textit{in-silico} with different 
degrees of cross-linking. To investigate how a regular and a disordered
distribution of the network's nodes influences the internal resistance to deformation, 
molecular dynamics simulations have been performed with explicit solvent 
molecules and with excluded volume for all the interactions. 
At the interface between two inmiscible liquids, nanogels adopt 
a shape that depends on the amount of nodes in their network 
and the interfacial tension of the confining medium. 

The nanogel permeability was analyzed by making use of a grid representation that picks 
the exact number of solvent particles inside deformable objects. With such an unbiased approach, it 
has been possible to obtain reliable density profiles for liquids inside 
such porous nanocavities. Comparisons were provided with respect to the case of 
a particle immersed in a single liquid. Nanogel's permeability is intrinsically 
related to the particle deformability. For a specific fraction of cross-links, 
the uptake of solvent particles in ideal nanogels is significantly bigger than their 
realistic counterpart due to its better packing efficiency. As the nanogels 
become less deformable, their uptake capacity reaches a plateau that  
depends on the interfacial forces between both liquids. Overall, the solvent 
uptake is optimized for cross-linking density $\sim 5$\% at soft interfaces 
and for $\sim 15-20\%$ at moderate and stiff interfaces.

We did not observe a fully mixed state for any of the interfacial strengths and degrees of cross-linking investigated, not even at soft interfaces. This is contrary 
to what is  reported via mean field 
approximations~\cite{Rumyantsev_2016_Potemkin}, or even with mesoscopic DPD 
simulations~\cite{Gumerov_2016_Potemkim,
Gumerov_2019_Potemkim}. Therefore our results suggest the need of incorporating  the
excluded volume for all the interactions (liquid-liquid, monomer-monomer and cross-interactions)
to investigate liquid miscibility within soft nanoparticles.
The results reported in this study indicate that miscibility is largely enhanced 
inside the nanocavities with respect to the bulk. In particular, we find that the 
mixing quality is enhanced by a factor between 2 and 5, depending on the degree 
of cross-linking and the interfacial strength. Most importantly, the specific 
network topology  only influences significantly the mixing 
quality for rigid interfaces, the disordered network providing a better mixing.

The emerging scenario proposes general guidelines, not only for setting the shape, uptake and mixing capacity of nanogels adsorbed at liquid-liquid interfaces,
but also for tuning the invasive capacity, which is an essential element to control
the catalysis rates in interfacial soft nanoreactors.
Furthermore, the results reported here should motivate future studies to unravel the specific role of the nanogel's instrinsic permeability within armored and bridging droplets. Work in this direction in progress.

\begin{acknowledgement}
DJAA acknowledges the career funding support from the DIPC foundation. 
This work has been supported by the projects PGC2018-094548-B-I00 (MCIU/AEI/FEDER, UE) 
and IT-1175-19 (Basque Government, Spain).
\end{acknowledgement}

\begin{suppinfo}
The following files are included to support the findings of this study
\begin{itemize}
 \item S1 (Figure): Overview of the computational workflow
 \item S2 (Tables): Average shape parameters of nanogels under 
different environments. 
 \item S3 (Figure): Swelling curves of nanogels in the perpendicular and 
parallel direction to the interface.
 \item S4 (Figure): Snapshot of solvent particles residing inside 
nanogels.
 \item S5-S6 (Videos): Side and top view videos for liquids inside nanogels.
 \item S7 (Figure): Density profiles of liquids inside nanogels with 7.5 and 
19.9 cross-links fractions.
\end{itemize}
\end{suppinfo}


\begin{mcitethebibliography}{59}
\providecommand*\natexlab[1]{#1}
\providecommand*\mciteSetBstSublistMode[1]{}
\providecommand*\mciteSetBstMaxWidthForm[2]{}
\providecommand*\mciteBstWouldAddEndPuncttrue
  {\def\EndOfBibitem{\unskip.}}
\providecommand*\mciteBstWouldAddEndPunctfalse
  {\let\EndOfBibitem\relax}
\providecommand*\mciteSetBstMidEndSepPunct[3]{}
\providecommand*\mciteSetBstSublistLabelBeginEnd[3]{}
\providecommand*\EndOfBibitem{}
\mciteSetBstSublistMode{f}
\mciteSetBstMaxWidthForm{subitem}{(\alph{mcitesubitemcount})}
\mciteSetBstSublistLabelBeginEnd
  {\mcitemaxwidthsubitemform\space}
  {\relax}
  {\relax}

\bibitem[Karg \latin{et~al.}(2019)Karg, Pich, Hellweg, Hoare, Lyon, Crassous,
  Suzuki, Gumerov, Schneider, Potemkin, and Richtering]{Karg_2019_Richtering}
Karg,~M.; Pich,~A.; Hellweg,~T.; Hoare,~T.; Lyon,~L.~A.; Crassous,~J.~J.;
  Suzuki,~D.; Gumerov,~R.~A.; Schneider,~S.; Potemkin,~I.~I.; Richtering,~W.
  Nanogels and Microgels: From Model Colloids to Applications, Recent
  Developments, and Future Trends. \emph{Langmuir} \textbf{2019}, \emph{35},
  6231--6255\relax
\mciteBstWouldAddEndPuncttrue
\mciteSetBstMidEndSepPunct{\mcitedefaultmidpunct}
{\mcitedefaultendpunct}{\mcitedefaultseppunct}\relax
\EndOfBibitem
\bibitem[Gupta \latin{et~al.}(2015)Gupta, Camargo, Stellbrink, Allgaier,
  Radulescu, Lindner, Zaccarelli, Likos, and Richter]{Gupta_2015_Richter}
Gupta,~S.; Camargo,~M.; Stellbrink,~J.; Allgaier,~J.; Radulescu,~A.;
  Lindner,~P.; Zaccarelli,~E.; Likos,~C.~N.; Richter,~D. Dynamic phase diagram
  of soft nanocolloids. \emph{Nanoscale} \textbf{2015}, \emph{7},
  13924--13934\relax
\mciteBstWouldAddEndPuncttrue
\mciteSetBstMidEndSepPunct{\mcitedefaultmidpunct}
{\mcitedefaultendpunct}{\mcitedefaultseppunct}\relax
\EndOfBibitem
\bibitem[Stuart \latin{et~al.}(2010)Stuart, Huck, Genzer, M{\"u}ller, Ober,
  Stamm, Sukhorukov, Szleifer, Tsukruk, Urban, Winnik, Zauscher, Luzinov, and
  Minko]{Stuart_2010_Minko}
Stuart,~M. A.~C.; Huck,~W.~T.; Genzer,~J.; M{\"u}ller,~M.; Ober,~C.; Stamm,~M.;
  Sukhorukov,~G.~B.; Szleifer,~I.; Tsukruk,~V.~V.; Urban,~M.; Winnik,~F.;
  Zauscher,~S.; Luzinov,~I.; Minko,~S. Emerging applications of
  stimuli-responsive polymer materials. \emph{Nature Materials} \textbf{2010},
  \emph{9}, 101--113\relax
\mciteBstWouldAddEndPuncttrue
\mciteSetBstMidEndSepPunct{\mcitedefaultmidpunct}
{\mcitedefaultendpunct}{\mcitedefaultseppunct}\relax
\EndOfBibitem
\bibitem[Plamper and Richtering(2017)Plamper, and
  Richtering]{Plamper_2017_Richtering}
Plamper,~F.~A.; Richtering,~W. Functional microgels and microgel systems.
  \emph{Accounts of Chemical Research} \textbf{2017}, \emph{50}, 131--140\relax
\mciteBstWouldAddEndPuncttrue
\mciteSetBstMidEndSepPunct{\mcitedefaultmidpunct}
{\mcitedefaultendpunct}{\mcitedefaultseppunct}\relax
\EndOfBibitem
\bibitem[Agrawal and Agrawal(2018)Agrawal, and Agrawal]{Agrawal_2018_Agrawal}
Agrawal,~G.; Agrawal,~R. Stimuli-responsive microgels and microgel-based
  systems: Advances in the exploitation of microgel colloidal properties and
  their interfacial activity. \emph{Polymers} \textbf{2018}, \emph{10},
  418\relax
\mciteBstWouldAddEndPuncttrue
\mciteSetBstMidEndSepPunct{\mcitedefaultmidpunct}
{\mcitedefaultendpunct}{\mcitedefaultseppunct}\relax
\EndOfBibitem
\bibitem[Echeverria \latin{et~al.}(2018)Echeverria, Fernandes, Godinho, Borges,
  and Soares]{Echeverria_2018_Soares}
Echeverria,~C.; Fernandes,~S.~N.; Godinho,~M.~H.; Borges,~J.~P.; Soares,~P.~I.
  Functional stimuli-responsive gels: Hydrogels and microgels. \emph{Gels}
  \textbf{2018}, \emph{4}, 54\relax
\mciteBstWouldAddEndPuncttrue
\mciteSetBstMidEndSepPunct{\mcitedefaultmidpunct}
{\mcitedefaultendpunct}{\mcitedefaultseppunct}\relax
\EndOfBibitem
\bibitem[Thorne \latin{et~al.}(2011)Thorne, Vine, and
  Snowden]{Thorne_2011_Snowden}
Thorne,~J.~B.; Vine,~G.~J.; Snowden,~M.~J. Microgel applications and commercial
  considerations. \emph{Colloid and Polymer Science} \textbf{2011}, \emph{289},
  625\relax
\mciteBstWouldAddEndPuncttrue
\mciteSetBstMidEndSepPunct{\mcitedefaultmidpunct}
{\mcitedefaultendpunct}{\mcitedefaultseppunct}\relax
\EndOfBibitem
\bibitem[Deshmukh \latin{et~al.}(2015)Deshmukh, van~den Ende, Stuart, Mugele,
  and Duits]{Deshmukh_2015_Duits}
Deshmukh,~O.~S.; van~den Ende,~D.; Stuart,~M.~C.; Mugele,~F.; Duits,~M.~H. Hard
  and soft colloids at fluid interfaces: Adsorption, interactions, assembly \&
  rheology. \emph{Advances in Colloid and Interface Science} \textbf{2015},
  \emph{222}, 215--227\relax
\mciteBstWouldAddEndPuncttrue
\mciteSetBstMidEndSepPunct{\mcitedefaultmidpunct}
{\mcitedefaultendpunct}{\mcitedefaultseppunct}\relax
\EndOfBibitem
\bibitem[Serpe(2019)]{Serpe_2019}
Serpe,~M.~J. Fine-tuned gel particles enable smart windows for energy
  efficiency. 2019\relax
\mciteBstWouldAddEndPuncttrue
\mciteSetBstMidEndSepPunct{\mcitedefaultmidpunct}
{\mcitedefaultendpunct}{\mcitedefaultseppunct}\relax
\EndOfBibitem
\bibitem[Torres \latin{et~al.}(2018)Torres, Andablo-Reyes, Murray, and
  Sarkar]{Torres_2018_Sarkar}
Torres,~O.; Andablo-Reyes,~E.; Murray,~B.~S.; Sarkar,~A. Emulsion microgel
  particles as high-performance bio-lubricants. \emph{ACS Applied Materials \&
  Interfaces} \textbf{2018}, \emph{10}, 26893--26905\relax
\mciteBstWouldAddEndPuncttrue
\mciteSetBstMidEndSepPunct{\mcitedefaultmidpunct}
{\mcitedefaultendpunct}{\mcitedefaultseppunct}\relax
\EndOfBibitem
\bibitem[Schmitt and Ravaine(2013)Schmitt, and Ravaine]{Schmitt_2013_Ravaine}
Schmitt,~V.; Ravaine,~V. Surface compaction versus stretching in Pickering
  emulsions stabilised by microgels. \emph{Current Opinion in Colloid \&
  Interface Science} \textbf{2013}, \emph{18}, 532--541\relax
\mciteBstWouldAddEndPuncttrue
\mciteSetBstMidEndSepPunct{\mcitedefaultmidpunct}
{\mcitedefaultendpunct}{\mcitedefaultseppunct}\relax
\EndOfBibitem
\bibitem[Yunker \latin{et~al.}(2014)Yunker, Chen, Gratale, Lohr, Still, and
  Yodh]{Yunker_2014_Yodh}
Yunker,~P.~J.; Chen,~K.; Gratale,~M.~D.; Lohr,~M.~A.; Still,~T.; Yodh,~A.
  Physics in ordered and disordered colloidal matter composed of poly
  (N-isopropylacrylamide) microgel particles. \emph{Reports on Progress in
  Physics} \textbf{2014}, \emph{77}, 056601\relax
\mciteBstWouldAddEndPuncttrue
\mciteSetBstMidEndSepPunct{\mcitedefaultmidpunct}
{\mcitedefaultendpunct}{\mcitedefaultseppunct}\relax
\EndOfBibitem
\bibitem[Kwok \latin{et~al.}(2019)Kwok, Sun, and Ngai]{Kwok_2019_Ngai}
Kwok,~M.-h.; Sun,~G.; Ngai,~T. Microgel Particles at Interfaces: Phenomena,
  Principles, and Opportunities in Food Sciences. \emph{Langmuir}
  \textbf{2019}, \emph{35}, 4205--4217\relax
\mciteBstWouldAddEndPuncttrue
\mciteSetBstMidEndSepPunct{\mcitedefaultmidpunct}
{\mcitedefaultendpunct}{\mcitedefaultseppunct}\relax
\EndOfBibitem
\bibitem[Murray(2019)]{Murray_2019}
Murray,~B.~S. Microgels at fluid-fluid interfaces for food and drinks.
  \emph{Advances in Colloid and Interface Science} \textbf{2019}, 101990\relax
\mciteBstWouldAddEndPuncttrue
\mciteSetBstMidEndSepPunct{\mcitedefaultmidpunct}
{\mcitedefaultendpunct}{\mcitedefaultseppunct}\relax
\EndOfBibitem
\bibitem[Ngai and Bon(2014)Ngai, and Bon]{Ngai_2014_Bon}
Ngai,~T.; Bon,~S.~A. \emph{Particle-stabilized emulsions and colloids:
  formation and applications}; Royal Society of Chemistry, 2014\relax
\mciteBstWouldAddEndPuncttrue
\mciteSetBstMidEndSepPunct{\mcitedefaultmidpunct}
{\mcitedefaultendpunct}{\mcitedefaultseppunct}\relax
\EndOfBibitem
\bibitem[Gupta \latin{et~al.}(2016)Gupta, Eral, Hatton, and
  Doyle]{Gupta_2016_Doyle}
Gupta,~A.; Eral,~H.~B.; Hatton,~T.~A.; Doyle,~P.~S. Nanoemulsions: formation{,}
  properties and applications. \emph{Soft Matter} \textbf{2016}, \emph{12},
  2826--2841\relax
\mciteBstWouldAddEndPuncttrue
\mciteSetBstMidEndSepPunct{\mcitedefaultmidpunct}
{\mcitedefaultendpunct}{\mcitedefaultseppunct}\relax
\EndOfBibitem
\bibitem[Wu \latin{et~al.}(2018)Wu, Lee, and Striolo]{Wu_2018_Striolo}
Wu,~N.; Lee,~D.; Striolo,~A. \emph{Anisotropic Particle Assemblies: Synthesis,
  Assembly, Modeling, and Applications}; Elsevier, 2018\relax
\mciteBstWouldAddEndPuncttrue
\mciteSetBstMidEndSepPunct{\mcitedefaultmidpunct}
{\mcitedefaultendpunct}{\mcitedefaultseppunct}\relax
\EndOfBibitem
\bibitem[Sicard \latin{et~al.}(2019)Sicard, Toro-Mendoza, and
  Striolo]{Sicard_2019_Striolo}
Sicard,~F.; Toro-Mendoza,~J.; Striolo,~A. Nanoparticles Actively Fragment
  Armored Droplets. \emph{ACS Nano} \textbf{2019}, \emph{13}, 9498--9503\relax
\mciteBstWouldAddEndPuncttrue
\mciteSetBstMidEndSepPunct{\mcitedefaultmidpunct}
{\mcitedefaultendpunct}{\mcitedefaultseppunct}\relax
\EndOfBibitem
\bibitem[Ngai \latin{et~al.}(2005)Ngai, Behrens, and
  Auweter]{Ngai_2005_Auweter}
Ngai,~T.; Behrens,~S.~H.; Auweter,~H. Novel emulsions stabilized by pH and
  temperature sensitive microgels. \emph{Chemical Communications}
  \textbf{2005}, 331--333\relax
\mciteBstWouldAddEndPuncttrue
\mciteSetBstMidEndSepPunct{\mcitedefaultmidpunct}
{\mcitedefaultendpunct}{\mcitedefaultseppunct}\relax
\EndOfBibitem
\bibitem[Wiese \latin{et~al.}(2016)Wiese, Tsvetkova, Daleiden, Spie{\ss}, and
  Richtering]{Wiese_2016_Richtering}
Wiese,~S.; Tsvetkova,~Y.; Daleiden,~N.~J.; Spie{\ss},~A.~C.; Richtering,~W.
  Microgel stabilized emulsions: Breaking on demand. \emph{Colloids and
  Surfaces A: Physicochemical and Engineering Aspects} \textbf{2016},
  \emph{495}, 193--199\relax
\mciteBstWouldAddEndPuncttrue
\mciteSetBstMidEndSepPunct{\mcitedefaultmidpunct}
{\mcitedefaultendpunct}{\mcitedefaultseppunct}\relax
\EndOfBibitem
\bibitem[Richtering and Saunders(2014)Richtering, and
  Saunders]{Richtering_2014_Saunders}
Richtering,~W.; Saunders,~B.~R. Gel architectures and their complexity.
  \emph{Soft Matter} \textbf{2014}, \emph{10}, 3695--3702\relax
\mciteBstWouldAddEndPuncttrue
\mciteSetBstMidEndSepPunct{\mcitedefaultmidpunct}
{\mcitedefaultendpunct}{\mcitedefaultseppunct}\relax
\EndOfBibitem
\bibitem[Siemes \latin{et~al.}(2018)Siemes, Nevskyi, Sysoiev, Turnhoff,
  Oppermann, Huhn, Richtering, and W{\"o}ll]{Siemes_2018_Richtering}
Siemes,~E.; Nevskyi,~O.; Sysoiev,~D.; Turnhoff,~S.~K.; Oppermann,~A.; Huhn,~T.;
  Richtering,~W.; W{\"o}ll,~D. Nanoscopic visualization of cross-linking
  density in polymer networks with diarylethene photoswitches. \emph{Angewandte
  Chemie International Edition} \textbf{2018}, \emph{57}, 12280--12284\relax
\mciteBstWouldAddEndPuncttrue
\mciteSetBstMidEndSepPunct{\mcitedefaultmidpunct}
{\mcitedefaultendpunct}{\mcitedefaultseppunct}\relax
\EndOfBibitem
\bibitem[Karanastasis \latin{et~al.}(2018)Karanastasis, Zhang, Kenath, Lessard,
  Bewersdorf, and Ullal]{Karanastasis_2018_Chaitanya}
Karanastasis,~A.~A.; Zhang,~Y.; Kenath,~G.~S.; Lessard,~M.~D.; Bewersdorf,~J.;
  Ullal,~C.~K. 3D mapping of nanoscale crosslink heterogeneities in microgels.
  \emph{Materials Horizons} \textbf{2018}, \emph{5}, 1130--1136\relax
\mciteBstWouldAddEndPuncttrue
\mciteSetBstMidEndSepPunct{\mcitedefaultmidpunct}
{\mcitedefaultendpunct}{\mcitedefaultseppunct}\relax
\EndOfBibitem
\bibitem[Geisel \latin{et~al.}(2012)Geisel, Isa, and
  Richtering]{Geisel_2012_Richtering}
Geisel,~K.; Isa,~L.; Richtering,~W. Unraveling the 3D localization and
  deformation of responsive microgels at oil/water interfaces: a step forward
  in understanding soft emulsion stabilizers. \emph{Langmuir} \textbf{2012},
  \emph{28}, 15770--15776\relax
\mciteBstWouldAddEndPuncttrue
\mciteSetBstMidEndSepPunct{\mcitedefaultmidpunct}
{\mcitedefaultendpunct}{\mcitedefaultseppunct}\relax
\EndOfBibitem
\bibitem[Cristofolini \latin{et~al.}(2018)Cristofolini, Orsi, and
  Isa]{Cristofolini_2018_Isa}
Cristofolini,~L.; Orsi,~D.; Isa,~L. Characterization of the dynamics of
  interfaces and of interface-dominated systems via spectroscopy and microscopy
  techniques. \emph{Current Opinion in Colloid \& Interface Science}
  \textbf{2018}, \emph{37}, 13--32\relax
\mciteBstWouldAddEndPuncttrue
\mciteSetBstMidEndSepPunct{\mcitedefaultmidpunct}
{\mcitedefaultendpunct}{\mcitedefaultseppunct}\relax
\EndOfBibitem
\bibitem[Mart{\'\i}n-Molina and Quesada-P{\'e}rez(2019)Mart{\'\i}n-Molina, and
  Quesada-P{\'e}rez]{Martin_2019_Quesada}
Mart{\'\i}n-Molina,~A.; Quesada-P{\'e}rez,~M. A review of coarse-grained
  simulations of nanogel and microgel particles. \emph{Journal of Molecular
  Liquids} \textbf{2019}, \emph{280}, 374--381\relax
\mciteBstWouldAddEndPuncttrue
\mciteSetBstMidEndSepPunct{\mcitedefaultmidpunct}
{\mcitedefaultendpunct}{\mcitedefaultseppunct}\relax
\EndOfBibitem
\bibitem[Kr{\"u}ger \latin{et~al.}(2013)Kr{\"u}ger, Frijters, G{\"u}nther,
  Kaoui, and Harting]{Kruger_2013_Harting}
Kr{\"u}ger,~T.; Frijters,~S.; G{\"u}nther,~F.; Kaoui,~B.; Harting,~J. Numerical
  simulations of complex fluid-fluid interface dynamics. \emph{The European
  Physical Journal Special Topics} \textbf{2013}, \emph{222}, 177--198\relax
\mciteBstWouldAddEndPuncttrue
\mciteSetBstMidEndSepPunct{\mcitedefaultmidpunct}
{\mcitedefaultendpunct}{\mcitedefaultseppunct}\relax
\EndOfBibitem
\bibitem[Camerin \latin{et~al.}(2018)Camerin, Gnan, Rovigatti, and
  Zaccarelli]{Camerin_2018_Zaccarelli}
Camerin,~F.; Gnan,~N.; Rovigatti,~L.; Zaccarelli,~E. Modelling realistic
  microgels in an explicit solvent. \emph{Scientific Reports} \textbf{2018},
  \emph{8}, 14426\relax
\mciteBstWouldAddEndPuncttrue
\mciteSetBstMidEndSepPunct{\mcitedefaultmidpunct}
{\mcitedefaultendpunct}{\mcitedefaultseppunct}\relax
\EndOfBibitem
\bibitem[Camerin \latin{et~al.}(2019)Camerin, Fern{\'a}ndez-Rodr{\'\i}guez,
  Rovigatti, Antonopoulou, Gnan, Ninarello, Isa, and
  Zaccarelli]{Camerin_2019_Zaccarelli}
Camerin,~F.; Fern{\'a}ndez-Rodr{\'\i}guez,~M.~{\'A}.; Rovigatti,~L.;
  Antonopoulou,~M.-N.; Gnan,~N.; Ninarello,~A.; Isa,~L.; Zaccarelli,~E.
  Microgels Adsorbed at Liquid--Liquid Interfaces: A Joint Numerical and
  Experimental Study. \emph{ACS nano} \textbf{2019}, \emph{13},
  4548--4559\relax
\mciteBstWouldAddEndPuncttrue
\mciteSetBstMidEndSepPunct{\mcitedefaultmidpunct}
{\mcitedefaultendpunct}{\mcitedefaultseppunct}\relax
\EndOfBibitem
\bibitem[Rumyantsev \latin{et~al.}(2016)Rumyantsev, Gumerov, and
  Potemkin]{Rumyantsev_2016_Potemkin}
Rumyantsev,~A.~M.; Gumerov,~R.~A.; Potemkin,~I.~I. A polymer microgel at a
  liquid--liquid interface: theory vs. computer simulations. \emph{Soft Matter}
  \textbf{2016}, \emph{12}, 6799--6811\relax
\mciteBstWouldAddEndPuncttrue
\mciteSetBstMidEndSepPunct{\mcitedefaultmidpunct}
{\mcitedefaultendpunct}{\mcitedefaultseppunct}\relax
\EndOfBibitem
\bibitem[Gumerov \latin{et~al.}(2016)Gumerov, Rumyantsev, Rudov, Pich,
  Richtering, Möller, and Potemkin]{Gumerov_2016_Potemkim}
Gumerov,~R.~A.; Rumyantsev,~A.~M.; Rudov,~A.~A.; Pich,~A.; Richtering,~W.;
  Moller,~M.; Potemkin,~I.~I. Mixing of two immiscible liquids within the
  polymer microgel adsorbed at their interface. \emph{ACS Macro Letters}
  \textbf{2016}, \emph{5}, 612--616\relax
\mciteBstWouldAddEndPuncttrue
\mciteSetBstMidEndSepPunct{\mcitedefaultmidpunct}
{\mcitedefaultendpunct}{\mcitedefaultseppunct}\relax
\EndOfBibitem
\bibitem[Gumerov \latin{et~al.}(2017)Gumerov, Rudov, Richtering, Möller, and
  Potemkin]{Gumerov_2017_Potenkim}
Gumerov,~R.~A.; Rudov,~A.~A.; Richtering,~W.; Moller,~M.; Potemkin,~I.~I.
  Amphiphilic arborescent copolymers and microgels: from unimolecular micelles
  in a selective solvent to the stable monolayers of variable density and
  nanostructure at a liquid Interface. \emph{ACS applied materials \&
  interfaces} \textbf{2017}, \emph{9}, 31302--31316\relax
\mciteBstWouldAddEndPuncttrue
\mciteSetBstMidEndSepPunct{\mcitedefaultmidpunct}
{\mcitedefaultendpunct}{\mcitedefaultseppunct}\relax
\EndOfBibitem
\bibitem[Gumerov \latin{et~al.}(2019)Gumerov, Filippov, Richtering, Pich, and
  Potemkin]{Gumerov_2019_Potemkim}
Gumerov,~R.~A.; Filippov,~S.~A.; Richtering,~W.; Pich,~A.; Potemkin,~I.~I.
  Amphiphilic microgels adsorbed at oil--water interfaces as mixers of two
  immiscible liquids. \emph{Soft matter} \textbf{2019}, \emph{15},
  3978--3986\relax
\mciteBstWouldAddEndPuncttrue
\mciteSetBstMidEndSepPunct{\mcitedefaultmidpunct}
{\mcitedefaultendpunct}{\mcitedefaultseppunct}\relax
\EndOfBibitem
\bibitem[Moreno and Lo~Verso(2018)Moreno, and Lo~Verso]{Moreno_2018_Loverso}
Moreno,~A.~J.; Lo~Verso,~F. Computational investigation of microgels: synthesis
  and effect of the microstructure on the deswelling behavior. \emph{Soft
  Matter} \textbf{2018}, \emph{14}, 7083--7096\relax
\mciteBstWouldAddEndPuncttrue
\mciteSetBstMidEndSepPunct{\mcitedefaultmidpunct}
{\mcitedefaultendpunct}{\mcitedefaultseppunct}\relax
\EndOfBibitem
\bibitem[Gnan \latin{et~al.}(2017)Gnan, Rovigatti, Bergman, and
  Zaccarelli]{Gnan_2017_Zaccarelli}
Gnan,~N.; Rovigatti,~L.; Bergman,~M.; Zaccarelli,~E. In silico synthesis of
  microgel particles. \emph{Macromolecules} \textbf{2017}, \emph{50},
  8777--8786\relax
\mciteBstWouldAddEndPuncttrue
\mciteSetBstMidEndSepPunct{\mcitedefaultmidpunct}
{\mcitedefaultendpunct}{\mcitedefaultseppunct}\relax
\EndOfBibitem
\bibitem[Minina \latin{et~al.}(2019)Minina, S{\'a}nchez, Likos, and
  Kantorovich]{Minina_2019_Kantorovich}
Minina,~E.~S.; S{\'a}nchez,~P.~A.; Likos,~C.~N.; Kantorovich,~S.~S. Studying
  synthesis confinement effects on the internal structure of nanogels in
  computer simulations. \emph{Journal of Molecular Liquids} \textbf{2019},
  \emph{289}, 111066\relax
\mciteBstWouldAddEndPuncttrue
\mciteSetBstMidEndSepPunct{\mcitedefaultmidpunct}
{\mcitedefaultendpunct}{\mcitedefaultseppunct}\relax
\EndOfBibitem
\bibitem[Ninarello \latin{et~al.}(2019)Ninarello, Crassous, Paloli, Camerin,
  Gnan, Rovigatti, Schurtenberger, and Zaccarelli]{Ninarello_2019}
Ninarello,~A.; Crassous,~J.~J.; Paloli,~D.; Camerin,~F.; Gnan,~N.;
  Rovigatti,~L.; Schurtenberger,~P.; Zaccarelli,~E. Modeling Microgels with a
  Controlled Structure across the Volume Phase Transition.
  \emph{Macromolecules} \textbf{2019}, \emph{52}, 7584--7592\relax
\mciteBstWouldAddEndPuncttrue
\mciteSetBstMidEndSepPunct{\mcitedefaultmidpunct}
{\mcitedefaultendpunct}{\mcitedefaultseppunct}\relax
\EndOfBibitem
\bibitem[Rudyak \latin{et~al.}(2019)Rudyak, Kozhunova, and
  Chertovich]{Rudyak_2019}
Rudyak,~V.~Y.; Kozhunova,~E.~Y.; Chertovich,~A.~V. Towards the realistic
  computer model of precipitation polymerization microgels. \emph{Scientific
  Reports} \textbf{2019}, \emph{9}, 13052\relax
\mciteBstWouldAddEndPuncttrue
\mciteSetBstMidEndSepPunct{\mcitedefaultmidpunct}
{\mcitedefaultendpunct}{\mcitedefaultseppunct}\relax
\EndOfBibitem
\bibitem[Rovigatti \latin{et~al.}(2019)Rovigatti, Gnan, Tavagnacco, Moreno, and
  Zaccarelli]{Rovigatti_2019_Zaccarelli}
Rovigatti,~L.; Gnan,~N.; Tavagnacco,~L.; Moreno,~A.~J.; Zaccarelli,~E.
  Numerical modelling of non-ionic microgels: an overview. \emph{Soft matter}
  \textbf{2019}, \emph{15}, 1108--1119\relax
\mciteBstWouldAddEndPuncttrue
\mciteSetBstMidEndSepPunct{\mcitedefaultmidpunct}
{\mcitedefaultendpunct}{\mcitedefaultseppunct}\relax
\EndOfBibitem
\bibitem[Ballard \latin{et~al.}(2019)Ballard, Law, and Bon]{Ballard_2019_Bon}
Ballard,~N.; Law,~A.~D.; Bon,~S.~A. Colloidal particles at fluid interfaces:
  behaviour of isolated particles. \emph{Soft matter} \textbf{2019}, \emph{15},
  1186--1199\relax
\mciteBstWouldAddEndPuncttrue
\mciteSetBstMidEndSepPunct{\mcitedefaultmidpunct}
{\mcitedefaultendpunct}{\mcitedefaultseppunct}\relax
\EndOfBibitem
\bibitem[{Lo Verso} \latin{et~al.}(2015){Lo Verso}, Pomposo, Colmenero, and
  Moreno]{Loverso_2015_Moreno}
{Lo Verso},~F.; Pomposo,~J.; Colmenero,~J.; Moreno,~A. {Simulation guided
  design of globular single-chain nanoparticles by tuning the solvent quality}.
  \emph{Soft Matter} \textbf{2015}, \emph{11}, 1369--1375\relax
\mciteBstWouldAddEndPuncttrue
\mciteSetBstMidEndSepPunct{\mcitedefaultmidpunct}
{\mcitedefaultendpunct}{\mcitedefaultseppunct}\relax
\EndOfBibitem
\bibitem[Pomposo(2017)]{Pomposo_2017}
Pomposo,~J. \emph{Single-Chain Polymer Nanoparticles: Synthesis,
  Characterization, Simulations, and Applications}; Wiley, 2017\relax
\mciteBstWouldAddEndPuncttrue
\mciteSetBstMidEndSepPunct{\mcitedefaultmidpunct}
{\mcitedefaultendpunct}{\mcitedefaultseppunct}\relax
\EndOfBibitem
\bibitem[Allen and Tildesley(2017)Allen, and Tildesley]{AllenBook_2017}
Allen,~M.~P.; Tildesley,~D.~J. \emph{Computer simulation of liquids}; Oxford
  University Press, 2017\relax
\mciteBstWouldAddEndPuncttrue
\mciteSetBstMidEndSepPunct{\mcitedefaultmidpunct}
{\mcitedefaultendpunct}{\mcitedefaultseppunct}\relax
\EndOfBibitem
\bibitem[Martinez \latin{et~al.}(2009)Martinez, Andrade, Birgin, and
  Martinez]{Packmol_2009}
Martinez,~L.; Andrade,~R.; Birgin,~E.~G.; Martinez,~J.~M. PACKMOL: A package
  for building initial configurations for molecular dynamics simulations.
  \emph{Journal of Computational Chemistry} \textbf{2009}, \emph{30},
  2157--2164\relax
\mciteBstWouldAddEndPuncttrue
\mciteSetBstMidEndSepPunct{\mcitedefaultmidpunct}
{\mcitedefaultendpunct}{\mcitedefaultseppunct}\relax
\EndOfBibitem
\bibitem[Abraham \latin{et~al.}(2015)Abraham, Murtola, Schulz, Pall, Smith,
  Hess, and Lindahl]{Gromacs_2015}
Abraham,~M.~J.; Murtola,~T.; Schulz,~R.; Pall,~S.; Smith,~J.~C.; Hess,~B.;
  Lindahl,~E. {GROMACS: High performance molecular simulations through
  multi-level parallelism from laptops to supercomputers}. \emph{SoftwareX}
  \textbf{2015}, \emph{1--2}, 19--25\relax
\mciteBstWouldAddEndPuncttrue
\mciteSetBstMidEndSepPunct{\mcitedefaultmidpunct}
{\mcitedefaultendpunct}{\mcitedefaultseppunct}\relax
\EndOfBibitem
\bibitem[Shuichi(1984)]{Nose_1984}
Shuichi,~N. A molecular dynamics method for simulations in the canonical
  ensemble. \emph{Molecular Physics} \textbf{1984}, \emph{52}, 255--268\relax
\mciteBstWouldAddEndPuncttrue
\mciteSetBstMidEndSepPunct{\mcitedefaultmidpunct}
{\mcitedefaultendpunct}{\mcitedefaultseppunct}\relax
\EndOfBibitem
\bibitem[Hoover(1985)]{Hoover_1985}
Hoover,~W.~G. Canonical dynamics: Equilibrium phase-space distributions.
  \emph{Phys. Rev. A} \textbf{1985}, \emph{31}, 1695--1697\relax
\mciteBstWouldAddEndPuncttrue
\mciteSetBstMidEndSepPunct{\mcitedefaultmidpunct}
{\mcitedefaultendpunct}{\mcitedefaultseppunct}\relax
\EndOfBibitem
\bibitem[Kremer and Grest(1990)Kremer, and Grest]{Kremer_1990_Grest}
Kremer,~K.; Grest,~G.~S. Dynamics of entangled linear polymer melts: A
  molecular-dynamics simulation. \emph{{J}. {C}hem. {P}hys.} \textbf{1990},
  \emph{92}, 5057--5086\relax
\mciteBstWouldAddEndPuncttrue
\mciteSetBstMidEndSepPunct{\mcitedefaultmidpunct}
{\mcitedefaultendpunct}{\mcitedefaultseppunct}\relax
\EndOfBibitem
\bibitem[Weeks \latin{et~al.}(1971)Weeks, Chandler, and
  Andersen]{Weeks_1971_Andersen}
Weeks,~J.~D.; Chandler,~D.; Andersen,~H.~C. Role of Repulsive Forces in
  Determining the Equilibrium Structure of Simple Liquids. \emph{The Journal of
  Chemical Physics} \textbf{1971}, \emph{54}, 5237--5247\relax
\mciteBstWouldAddEndPuncttrue
\mciteSetBstMidEndSepPunct{\mcitedefaultmidpunct}
{\mcitedefaultendpunct}{\mcitedefaultseppunct}\relax
\EndOfBibitem
\bibitem[Rawdon \latin{et~al.}(2008)Rawdon, Kern, Piatek, Plunkett, Stasiak,
  and Millett]{Rawdon_2008_Millett}
Rawdon,~E.~J.; Kern,~J.~C.; Piatek,~M.; Plunkett,~P.; Stasiak,~A.;
  Millett,~K.~C. Effect of knotting on the shape of polymers.
  \emph{Macromolecules} \textbf{2008}, \emph{41}, 8281--8287\relax
\mciteBstWouldAddEndPuncttrue
\mciteSetBstMidEndSepPunct{\mcitedefaultmidpunct}
{\mcitedefaultendpunct}{\mcitedefaultseppunct}\relax
\EndOfBibitem
\bibitem[Rudnick and Gaspari(1987)Rudnick, and Gaspari]{Rudnick_1987_Gaspari}
Rudnick,~J.; Gaspari,~G. The shapes of random walks. \emph{Science}
  \textbf{1987}, \emph{237}, 384--389\relax
\mciteBstWouldAddEndPuncttrue
\mciteSetBstMidEndSepPunct{\mcitedefaultmidpunct}
{\mcitedefaultendpunct}{\mcitedefaultseppunct}\relax
\EndOfBibitem
\bibitem[Destribats \latin{et~al.}(2014)Destribats, Eyharts, Lapeyre, Sellier,
  Varga, Ravaine, and Schmitt]{Destribats_2014_Schmitt}
Destribats,~M.; Eyharts,~M.; Lapeyre,~V.; Sellier,~E.; Varga,~I.; Ravaine,~V.;
  Schmitt,~V. Impact of pNIPAM microgel size on its ability to stabilize
  Pickering emulsions. \emph{Langmuir} \textbf{2014}, \emph{30},
  1768--1777\relax
\mciteBstWouldAddEndPuncttrue
\mciteSetBstMidEndSepPunct{\mcitedefaultmidpunct}
{\mcitedefaultendpunct}{\mcitedefaultseppunct}\relax
\EndOfBibitem
\bibitem[Keal \latin{et~al.}(2017)Keal, Lapeyre, Ravaine, Schmitt, and
  Monteux]{Keal_2017_Monteux}
Keal,~L.; Lapeyre,~V.; Ravaine,~V.; Schmitt,~V.; Monteux,~C. Drainage dynamics
  of thin liquid foam films containing soft PNiPAM microgels: influence of the
  cross-linking density and concentration. \emph{Soft Matter} \textbf{2017},
  \emph{13}, 170--180\relax
\mciteBstWouldAddEndPuncttrue
\mciteSetBstMidEndSepPunct{\mcitedefaultmidpunct}
{\mcitedefaultendpunct}{\mcitedefaultseppunct}\relax
\EndOfBibitem
\bibitem[Tatry \latin{et~al.}(2019)Tatry, Laurichesse, Perro, Ravaine, and
  Schmitt]{Tatry_2019_Schmitt}
Tatry,~M.~C.; Laurichesse,~E.; Perro,~A.; Ravaine,~V.; Schmitt,~V. Kinetics of
  spontaneous microgels adsorption and stabilization of emulsions produced
  using microfluidics. \emph{Journal of colloid and interface science}
  \textbf{2019}, \emph{548}, 1--11\relax
\mciteBstWouldAddEndPuncttrue
\mciteSetBstMidEndSepPunct{\mcitedefaultmidpunct}
{\mcitedefaultendpunct}{\mcitedefaultseppunct}\relax
\EndOfBibitem
\bibitem[Destribats \latin{et~al.}(2011)Destribats, Lapeyre, Wolfs, Sellier,
  Leal-Calderon, Ravaine, and Schmitt]{Destribats_2011_Schmitt}
Destribats,~M.; Lapeyre,~V.; Wolfs,~M.; Sellier,~E.; Leal-Calderon,~F.;
  Ravaine,~V.; Schmitt,~V. Soft microgels as Pickering emulsion stabilisers:
  role of particle deformability. \emph{Soft Matter} \textbf{2011}, \emph{7},
  7689--7698\relax
\mciteBstWouldAddEndPuncttrue
\mciteSetBstMidEndSepPunct{\mcitedefaultmidpunct}
{\mcitedefaultendpunct}{\mcitedefaultseppunct}\relax
\EndOfBibitem
\bibitem[Destribats \latin{et~al.}(2013)Destribats, Wolfs, Pinaud, Lapeyre,
  Sellier, Schmitt, and Ravaine]{Destribats_2013_Ravaine}
Destribats,~M.; Wolfs,~M.; Pinaud,~F.; Lapeyre,~V.; Sellier,~E.; Schmitt,~V.;
  Ravaine,~V. Pickering emulsions stabilized by soft microgels: influence of
  the emulsification process on particle interfacial organization and emulsion
  properties. \emph{Langmuir} \textbf{2013}, \emph{29}, 12367--12374\relax
\mciteBstWouldAddEndPuncttrue
\mciteSetBstMidEndSepPunct{\mcitedefaultmidpunct}
{\mcitedefaultendpunct}{\mcitedefaultseppunct}\relax
\EndOfBibitem
\bibitem[Harrer \latin{et~al.}(2019)Harrer, Rey, Ciarella, Lowen, Janssen,
  and Vogel]{Harrer_2019_Vogel}
Harrer,~J.; Rey,~M.; Ciarella,~S.; Lowen,~H.; Janssen,~L.~M.; Vogel,~N.
  Stimuli-responsive behavior of PNiPAm microgels under interfacial
  confinement. \emph{Langmuir} \textbf{2019}, \emph{35}, 10512--10521\relax
\mciteBstWouldAddEndPuncttrue
\mciteSetBstMidEndSepPunct{\mcitedefaultmidpunct}
{\mcitedefaultendpunct}{\mcitedefaultseppunct}\relax
\EndOfBibitem
\bibitem[Maestro(2019)]{Maestro_2019}
Maestro,~A. Tailoring the interfacial assembly of colloidal particles by
  engineering the mechanical properties of the interface. \emph{Current opinion
  in colloid \& interface science} \textbf{2019}, \emph{39}, 232--250\relax
\mciteBstWouldAddEndPuncttrue
\mciteSetBstMidEndSepPunct{\mcitedefaultmidpunct}
{\mcitedefaultendpunct}{\mcitedefaultseppunct}\relax
\EndOfBibitem
\end{mcitethebibliography}

\providecommand{\latin}[1]{#1}
\providecommand*\mcitethebibliography{\thebibliography}
\csname @ifundefined\endcsname{endmcitethebibliography}
  {\let\endmcitethebibliography\endthebibliography}{}




\end{document}